\documentclass[twocolumn]{aastex63}

\usepackage{mathrsfs}

\shorttitle{JWST unveils obscured quasars at $z > 6$}
\shortauthors{Matsuoka et al.}

\begin{document}

\title{SHELLQs. Bridging the gap: JWST unveils obscured quasars in the most luminous galaxies at $z > 6$}

\correspondingauthor{Yoshiki Matsuoka}
\email{matsuoka.yoshiki.ld@ehime-u.ac.jp}


\author[0000-0001-5063-0340]{Yoshiki Matsuoka}
\affil{Research Center for Space and Cosmic Evolution, Ehime University, Matsuyama, Ehime 790-8577, Japan.}

\author[0000-0003-2984-6803]{Masafusa Onoue}
\affiliation{Waseda Institute for Advanced Study (WIAS), Waseda University, Shinjuku, Tokyo 169-0051, Japan.}
\affiliation{Kavli Institute for the Physics and Mathematics of the Universe, WPI, The University of Tokyo, Kashiwa, Chiba 277-8583, Japan.}
\affiliation{Center for Data-Driven Discovery, Kavli IPMU (WPI), UTIAS, The University of Tokyo, Kashiwa, Chiba 277-8583, Japan.}
\affiliation{Kavli Institute for Astronomy and Astrophysics, Peking University, Beijing 100871, P.R.China.}

\author[0000-0002-4923-3281]{Kazushi Iwasawa}
\affil{ICREA and Institut de Ci{\`e}ncies del Cosmos, Universitat de Barcelona, IEEC-UB, Mart{\'i} i Franqu{\`e}s, 1, 08028 Barcelona, Spain.}

\author[0000-0003-4569-1098]{Kentaro Aoki}
\affil{Subaru Telescope, National Astronomical Observatory of Japan, Hilo, HI 96720, USA.}

\author[0000-0002-0106-7755]{Michael A. Strauss}
\affil{Department of Astrophysical Sciences, Princeton University, Peyton Hall, Princeton, NJ 08544, USA.}

\author[0000-0002-0000-6977]{John D. Silverman}
\affiliation{Kavli Institute for the Physics and Mathematics of the Universe, WPI, The University of Tokyo, Kashiwa, Chiba 277-8583, Japan.}
\affiliation{Department of Astronomy, School of Science, The University of Tokyo, Tokyo 113-0033, Japan.}
\affiliation{Center for Data-Driven Discovery, Kavli IPMU (WPI), UTIAS, The University of Tokyo, Kashiwa, Chiba 277-8583, Japan}
\affiliation{Center for Astrophysical Sciences, Department of Physics \& Astronomy, Johns Hopkins University, Baltimore, MD 21218, USA}\

\author[0000-0001-8917-2148]{Xuheng Ding}
\affil{School of Physics and Technology, Wuhan University, Wuhan 430072, China.}
\affil{Kavli Institute for the Physics and Mathematics of the Universe, WPI, The University of Tokyo, Kashiwa, Chiba 277-8583, Japan.}

\author[0000-0002-2099-0254]{Camryn L. Phillips}
\affil{Department of Astrophysical Sciences, Princeton University, Peyton Hall, Princeton, NJ 08544, USA.}

\author[0000-0002-2651-1701]{Masayuki Akiyama}
\affil{Astronomical Institute, Tohoku University, Aoba, Sendai, 980-8578, Japan.}

\author[0009-0007-0864-7094]{Junya Arita}
\affiliation{Department of Astronomy, School of Science, The University of Tokyo, Tokyo 113-0033, Japan.}

\author[0000-0001-6186-8792]{Masatoshi Imanishi}
\affil{National Astronomical Observatory of Japan, Mitaka, Tokyo 181-8588, Japan.}
\affil{Department of Astronomical Science, Graduate University for Advanced Studies (SOKENDAI), Mitaka, Tokyo 181-8588, Japan.}

\author[0000-0001-9452-0813]{Takuma Izumi}
\affil{National Astronomical Observatory of Japan, Mitaka, Tokyo 181-8588, Japan.}

\author[0000-0003-3954-4219]{Nobunari Kashikawa}
\affil{Department of Astronomy, School of Science, The University of Tokyo, Tokyo 113-0033, Japan.}

\author[0000-0002-3866-9645]{Toshihiro Kawaguchi}
\affil{Graduate School of Science and Engineering, University of Toyama, Toyama, Toyama 930-8555, Japan.}

\author[0000-0003-3214-9128]{Satoshi Kikuta}
\affil{Department of Astronomy, School of Science, The University of Tokyo, Tokyo 113-0033, Japan.}

\author[0000-0002-4052-2394]{Kotaro Kohno}
\affil{Institute of Astronomy, The University of Tokyo, Mitaka, Tokyo 181-0015, Japan.}
\affil{Research Center for the Early Universe, University of Tokyo, Tokyo 113-0033, Japan.}

\author[0000-0003-1700-5740]{Chien-Hsiu Lee}
\affil{Hobby-Eberly Telescope, McDonald Observatory, Fort Davis, TX 79734, USA.}

\author[0000-0002-7402-5441]{Tohru Nagao}
\affil{Research Center for Space and Cosmic Evolution, Ehime University, Matsuyama, Ehime 790-8577, Japan.}

\author[0000-0003-3769-6630]{Ayumi Takahashi}
\affil{National Astronomical Observatory of Japan, Mitaka, Tokyo 181-8588, Japan.}

\author[0000-0002-3531-7863]{Yoshiki Toba}
\affiliation{Department of Physical Sciences, Ritsumeikan University, Kusatsu, Shiga 525-8577, Japan.}
\affiliation{National Astronomical Observatory of Japan, Mitaka, Tokyo 181-8588, Japan.}
\affiliation{Academia Sinica Institute of Astronomy and Astrophysics, Taipei 10617, Taiwan.}
\affiliation{Research Center for Space and Cosmic Evolution, Ehime University, Matsuyama, Ehime 790-8577, Japan.}



\begin{abstract}

The unprecedented sensitivity of the James Webb Space Telescope (JWST) has uncovered a surprisingly abundant population of mildly obscured, low-luminosity active galactic nuclei (AGNs) in the epoch of reionization (EoR).
However, the link between these objects and classical unobscured quasars remains a mystery.
Here we report the discovery of obscured quasars hosted by the most luminous galaxies at $z > 6$, possibly bridging the gap between the two AGN populations.
The 13 objects presented here were originally selected from a rest-frame ultraviolet (UV) imaging survey over  $>$1000 deg$^2$, 
and were known to have luminous ($>10^{43}$ erg s$^{-1}$) Ly$\alpha$ emission. 
With JWST/NIRSpec follow-up observations, 
we found that 7 out of 11 objects with narrow Ly$\alpha$ exhibit a broad component in \ion{H}{1} Balmer lines and \ion{He}{1} lines, but not in [\ion{O}{3}] and other forbidden lines.
Mild dust obscuration ($0 < A_V < 3$) is inferred from the Balmer decrements. 
The estimated intrinsic luminosities suggest that our broad line (BL) objects 
are the long-sought UV-obscured counterparts of luminous quasars in the EoR.
They host supermassive black holes (SMBHs) with masses $10^{7.8 - 9.1} M_\odot$, undergoing sub-Eddington to Eddington accretion.
Most of the BL objects are spatially unresolved, and are close to ``little red dots" with their blue rest-UV and red rest-optical colors.
We estimate the AGN number density among similarly luminous Ly$\alpha$ emitters to be larger than $2 \times 10^{-8}$ Mpc$^{-3}$.
This density is comparable to that of classical quasars with similar continuum luminosities, suggesting that a substantial fraction of active SMBHs are obscured in the EoR and have been overlooked in past rest-UV surveys.

\end{abstract}




\section{Introduction} \label{sec:intro}

The first few years of James Webb Space Telescope \citep[JWST;][]{rigby23} observations have transformed our understanding of the epoch of reionization (EoR, referring to $z > 6$ in this paper), when the first generations of stars, galaxies, and supermassive black holes (SMBHs) formed.
The unprecedented infrared (IR) sensitivity of the telescope has revealed that this cosmic epoch is surprisingly rich with luminous galaxies, with rest-frame ultraviolet (UV) absolute magnitudes of $M_{\rm UV} < -20$ \citep[e.g.,][]{castellano22, finkelstein23, labbe23}.
Galaxies are now known to exist out to $z > 14$ \citep{carniani24a, carniani24b, naidu25_gal}.
The number density of such early galaxies is significantly higher than predicted before JWST was launched, challenging our models for early galaxy evolution \citep[e.g.,][]{donnan24, robertson24}.

JWST has also found that an unexpectedly high fraction of galaxies have 
a broad component in their H$\alpha$ and H$\beta$ emission lines, mainly at $z > 4$
 \citep[e.g.,][]{carnall23, harikane23,  larson23, onoue23, kokorev23, ubler23, fujimoto24, furtak24, juodzbalis24, juodzbalis25,killi24,labbe24, lin24, lin25, kocevski23,kocevski24,  maiolino24, schindler24, taylor24, taylor25, tripodi24, wang24b, akins25, naidu25_lrd}.
 The observed broad line widths, with full width at half maximum (FWHM) exceeding 1000 km s$^{-1}$, are much larger than usually seen in normal star-forming galaxies.
The absence of a similar component in bright [\ion{O}{3}] $\lambda$5007 and other forbidden lines suggests that the broad lines are emitted from dense gas clouds \citep[e.g.,][]{greene24}.
The most common interpretation is that the gas clouds belong to active galactic nuclei (AGNs), whose luminosities ($M_{\rm UV} > -21$) are much lower than those of classical EoR quasars discovered 
over the past decades \citep[$M_{\rm UV} < -24$; e.g.,][and references therein]{fan23}.

A significant fraction of the broad H$\alpha$/H$\beta$ emitters (BHEs) are members of a galaxy population called ``little red dots (LRDs)", which were also newly identified by JWST \citep[e.g.,][]{matthee24}.
LRDs are distinguished by their compact morphology and characteristic V-shaped spectra with blue rest-UV and red rest-optical colors \citep[e.g.,][]{labbe23, labbe25, setton24}.
Analyses of their continuum slopes and Balmer decrements found evidence of mild dust obscuration, with up to several magnitudes of visual extinction $A_V$ \citep[e.g.,][]{casey24, greene24}.
Overall, the accumulated observations suggest that JWST has identified a new class of galaxies hosting mildly obscured low-luminosity AGNs, which manifest as BHEs and/or LRDs
\citep[see also, e.g.,][for cases with X-ray detection]{goulding23,bogdan24}.
The number density of such objects is at least 10 -- 100 times higher than predicted by the UV luminosity function of classical quasars extrapolated to the faint end
\citep[e.g.,][]{harikane23, greene24, kocevski24, maiolino24, ma25b}.

On the other hand, there are important ways in which BHEs/LRDs differ from classical quasars. 
The former objects are not detected in ultra-deep Chandra observations, indicating X-ray luminosities much lower than inferred from broad H$\alpha$/H$\beta$ fluxes \citep{ananna24, maiolino24_xray, yue24}.
Their rest-frame UV to optical spectra have been reported to be difficult to reproduce with a combination of the existing AGN and galaxy models \citep[][but see, e.g., \citet{labbe24}]{ma25}.
An absence of mid-IR flux excess rules out significant emission from hot dust, which usually is a feature of AGN spectral energy distribution \citep[SED; e.g.,][]{wang24,williams24,setton25}.
They also remain undetected in deep stacked data at radio frequencies, even though the sensitivities are high enough to detect emission from radio-quiet AGNs at similar redshifts \citep[e.g.,][but see \cite{gloudemans25}]{akins24, mazzolari24, perger25}.
Finally, no flux variability is detected from them, even though the existing JWST photometry provides sufficient time sampling and sensitivity for detecting variability from normal AGNs \citep{kokubo24, tee24}.
The unique features of BHEs/LRDs (assuming that they are a family of AGNs) may be caused by, e.g., super-Eddington accretion of matter onto supermassive black holes \citep[SMBHs;][]{inayoshi24, pacucci24}, but it remains challenging to explain all the observed properties, including the high number density, with existing AGN models.
Alternative scenarios have also been proposed, e.g., very high stellar density at the galaxy centers responsible for at least part of the BHEs \citep{baggen24}. 

\begin{figure*}
\epsscale{1}
\plotone{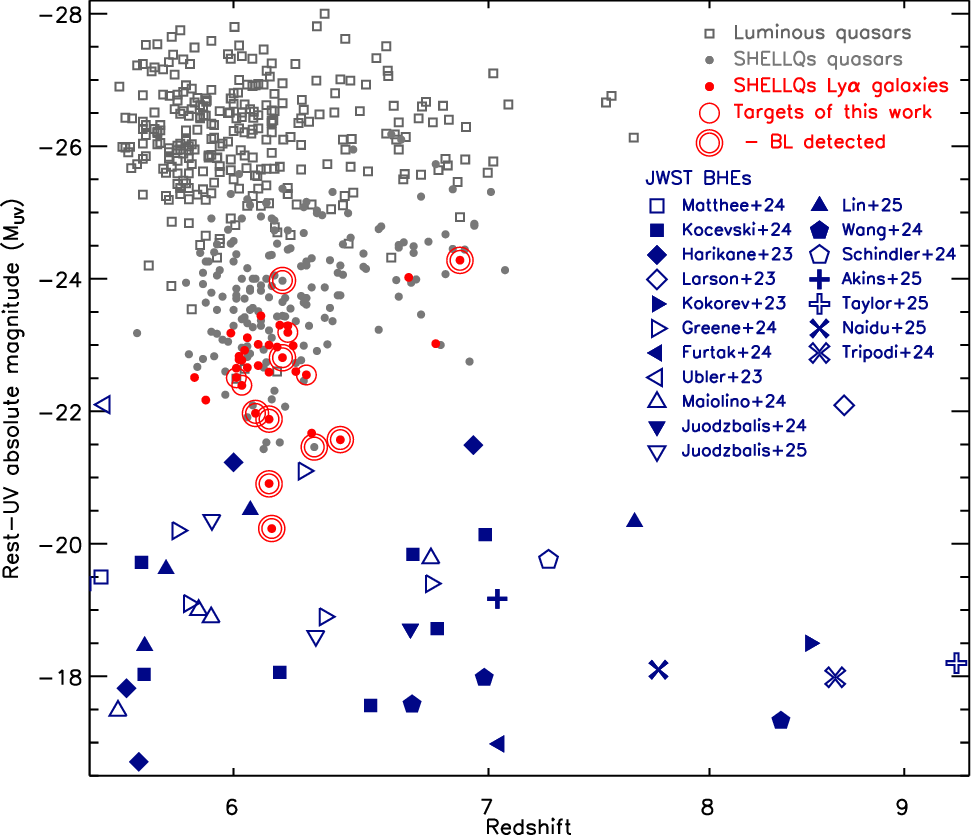}
\caption{
Redshifts and rest-UV absolute magnitudes of (mostly) luminous quasars in the literature (open squares), SHELLQs low-luminosity quasars (grey dots), SHELLQs galaxies with 
Ly$\alpha$ luminosity $L_{\rm Ly\alpha} > 10^{43}$ erg s$^{-1}$ (red dots), and JWST BHEs 
\citep[dark blue symbols;][]{harikane23,kokorev23,ubler23,furtak24,greene24,juodzbalis24,juodzbalis25,kocevski24,maiolino24,matthee24,schindler24,tripodi24,wang24b,akins25,lin25,naidu25_lrd,taylor25} at $z > 5.5$.
The 13 JWST targets of this work are marked by the open circles, and those with broad line detections (see text) are marked by the larger open circles.
The $M_{\rm UV}$ values are as measured, i.e., not corrected for dust extinction.
\label{fig:agn_rel}}
\end{figure*}

In summary, two (possible) AGN populations are now known in the EoR, and there is a clear gap between their multiple observed properties.
One population is classical UV-luminous ($M_{\rm UV} < -24$) quasars with little dust obscuration, and the other is mildly obscured low-luminosity ($M_{\rm UV} > -21$) AGNs, which only JWST can identify as BHEs and/or LRDs.
The former population features X-ray, mid-IR, and radio emission and flux variability, characteristic of lower-redshift AGNs, while the latter population shows none of these properties.
A key to bridging the gap may be hidden in the intermediate luminosity range ($-24 < M_{\rm UV} < -21$).
It is a relatively uncharted territory in the EoR, since objects with such luminosities are too rare on the sky to be found in deep pencil-beam surveys such as those carried out by JWST, 
but are also too faint to be detected in most of the past wide-field surveys.

The ``Subaru High-$z$ Exploration of Low-Luminosity Quasars" \citep[SHELLQs;][]{p1} project is uniquely suited for exploring the above key luminosity range.
As displayed in Figure \ref{fig:agn_rel}, the project has so far discovered $\sim$180 faint broad-line quasars at $5.6 < z < 7.1$, including objects whose luminosities are comparable to some of the BHEs (LRDs are included in the figure only when they have spectroscopic information and BHE signatures).
The multi-wavelength properties of the quasars have been extensively studied with Atacama Large Millimeter/submillimeter Array \citep[ALMA;][]{izumi18,izumi19,izumi21a,izumi21b,sawamura25}, JWST \citep[][M. Onoue et al., in prep.; C. Phillips et al., in prep.]{ding23,ding25,lyu24,onoue24,stone24}, 
and other facilities \citep{onoue19,onoue21,takahashi24}.
The SHELLQs quasars are characterized by little dust extinction except for a few reddened objects \citep{kato20, iwamoto25}, so they likely represent {\it the lowest-luminosity family of classical unobscured quasars}, at least as discovered thus far.

\begin{deluxetable*}{cccccc}
\tablecaption{Targets of this work\label{tab:targets}}
\tablewidth{0pt}
\tablehead{
\colhead{Name (Formal HSC name)} & \colhead{$z_{\rm Ly\alpha}$} & \colhead{$\log\ (L_{\rm Ly\alpha}/{\rm [erg\ s}^{-1}])$} 
& \colhead{FWHM$_{\rm Ly\alpha}$} & \colhead{$M_{\rm UV}$} & \colhead{Ref}\\
\colhead{} & \colhead{} & \colhead{} & \colhead{(km s$^{-1}$)} & \colhead{} & \colhead{}
}
\startdata
G01 ($J$142331.71$-$001809.1)   & 6.13  & 44.30 $\pm$ 0.01 &  $<$230 & $-$21.88  $\pm$ 0.20 & 2 \\ 
G02 ($J$084456.62$+$022640.5)   & 6.40 & 44.05  $\pm$ 0.01 & $<$230  &  $-$21.57  $\pm$ 0.47 & 4 \\ 
G03 ($J$093543.32$-$011033.3)   & 6.08 & 44.12  $\pm$ 0.01 & $<$230  &  $-$21.97  $\pm$ 0.18 & 4 \\ 
G04 ($J$125437.08$-$001410.7)   & 6.13  & 44.03  $\pm$ 0.01 & 310 $\pm$ 20  &  $-$20.91  $\pm$ 0.32 & 4\\  
G05 ($J$020719.59$+$023826.0)   & 6.14  & 44.06 $\pm$  0.01 & 400 $\pm$ 30  & $-$20.23  $\pm$ 1.59 & 5 \\ 
G06 ($J$142322.01$+$020612.8)   & 6.88  & 44.33  $\pm$ 0.02 &  $<$230 & $-$24.28  $\pm$ 0.34 & 5 \\ 
G07 ($J$223212.03$+$001238.4)   &  6.18 & 44.06  $\pm$ 0.01  & 300 $\pm$ 30  & $-$22.81  $\pm$ 0.10 & 1, 2 \\ 
G08 ($J$090544.65$+$030058.9)   & 6.27  & 43.89  $\pm$ 0.02 & 250 $\pm$ 40  &  $-$22.55  $\pm$ 0.11 & 2 \\ 
G09 ($J$141612.71$+$001546.2)   & 6.03 & 43.86  $\pm$ 0.01  & 230 $\pm$ 20  & $-$22.39  $\pm$ 0.10 & 2 \\ 
G10 ($J$085348.84$+$013911.0)   &  6.01  & 43.80  $\pm$ 0.01  &  $<$230  &  $-$22.51  $\pm$ 0.14 & 2 \\ 
G11 ($J$021721.59$-$020852.6)    & 6.20  & 43.33  $\pm$ 0.04 &  $<$230  &  $-$23.19  $\pm$ 0.04 & 2  \\ 
Q01 ($J$084408.61$-$013216.5)    & 6.18  & 44.25  $\pm$ 0.02  &  1610 $\pm$ 280 &  $-$23.97  $\pm$ 0.11 & 3 \\ 
Q02 ($J$114658.90$-$000537.6)    & 6.30  & 44.18  $\pm$ 0.02 & 330 $\pm$ 100  &  $-$21.46  $\pm$ 0.63 & 3 \\ 
\enddata
\tablecomments{
The Ly$\alpha$ redshifts ($z_{\rm Ly\alpha}$) were determined from the observed peak wavelength of Ly$\alpha$.
The Ly$\alpha$ luminosities ($L_{\rm Ly\alpha}$), FWHMs, and the rest-UV absolute magnitudes ($M_{\rm UV}$) are reported as measured, without correction for dust extinction or IGM absorption.
The last column lists the discovery papers; (1) \citet{p1}, (2) \citet{p2}, (3) \citet{p4}, (4) \citet{p10}, (5) Y. Matsuoka et al. (in prep.).
}
\end{deluxetable*}

SHELLQs has also identified 34 galaxies with strong Ly$\alpha$ emission at $5.8 < z < 6.9$ (the red dots in Figure \ref{fig:agn_rel}).
Based on the line widths (FWHM $<$ 500 km s$^{-1}$) and luminosities ($L_{\rm Ly\alpha} > 10^{43}$ erg s$^{-1}$), we suggested in our past publications \citep[e.g.,][]{p2} 
that these galaxies are 
candidate narrow-line quasars.
The present paper describes the rest-optical spectroscopy of 11 galaxies from this sample, plus two SHELLQs quasars, obtained by NIRSpec on JWST. 
We detected broad \ion{H}{1} and \ion{He}{1} lines and mild dust obscuration in the majority of the targets, which suggests that they represent 
{\it the highest-luminosity family of BHEs}. 
Some of them meet the definition of LRDs, while several others have colors close to the borderline between LRDs and non-LRDs.

This paper is structured as follows.
Section \S \ref{sec:obs} describes the sample, NIRSpec observations and data reduction, and spectral modeling.
The results and discussions appear in Section \S \ref{sec:results}, followed by a summary in  \S \ref{sec:summary}.
We adopt the cosmological parameters $H_0$ = 70 km s$^{-1}$ Mpc$^{-1}$, $\Omega_{\rm M}$ = 0.3, and $\Omega_{\rm \Lambda}$ = 0.7.
All magnitudes are presented in the AB system \citep{oke83}. 
The rest-UV absolute magnitudes ($M_{\rm UV}$) and Ly$\alpha$ line luminosities ($L_{\rm Ly\alpha}$) and widths are reported as measured, i.e., without correction for dust extinction or intergalactic medium (IGM) absorption, unless otherwise noted.
We report wavelengths in both the rest frame ($\lambda_{\rm rest}$) and the observed frame ($\lambda_{\rm obs}$), depending on the context.
A companion paper \citep{iwasawa25} reports the results from Chandra X-ray observations for a subset of the objects presented here.

\section{Observations} \label{sec:obs}

\subsection{Sample selection}

The sample presented here was originally selected from the Hyper Suprime-Cam \citep[HSC;][]{miyazaki18} Subaru Strategic Program (SSP) imaging survey.
The survey consisted of three layers named Wide, Deep, and UltraDeep, covering (1100, 26, 3.5) deg$^2$ respectively, with 5$\sigma$ limiting magnitude of $i_{\rm AB}$ = (25.9, 26.8, 27.4) for point sources \citep{aihara18}.
We launched the SHELLQs project to search for EoR quasars from this imaging dataset \citep{p1}.
Here we briefly describe how the actual quasar selection worked. 
We first selected objects from the three layers with the conditions ($z_{\rm AB} < 24.5$ $\cap$ $\sigma_z < 0.155$ $\cap$ $i_{\rm AB} - z_{\rm AB} > 1.5$) $\cup$ ($y_{\rm AB} < 25.0$ $\cap$ $\sigma_y < 0.217$ $\cap$ $z_{\rm AB} - y_{\rm AB} > 0.8$). 
The two color cuts were used to extract objects with strong IGM absorption expected at $z \sim 6$ and $z \sim 7$, respectively.
Sources with $g$- or $r$-band detection or with any critical quality flags were excluded.
We  matched the selected objects with public near-IR catalogs, and 
calculated a Bayesian probability for each candidate being a high-$z$ quasar rather than a Galactic brown dwarf, based on its broadband SED. 
Since our algorithm contains no galaxy models, the resultant quasar candidates include UV-bright galaxies at similar redshifts, due to their similar rest-UV spectral shapes. 
We screened the stacked and pre-stacked HSC images of every candidate, and removed false detections, transients, and other problematic sources at this stage.
Finally we performed optical spectroscopic follow-up of the candidates, using the Subaru Telescope and the Gran Telescopio Canarias (GTC).
Higher observing priorities were given to candidates with higher quasar probability, brighter magnitudes, and more point-like morphology.
The full details of the selection process are provided in our past publications \citep[e.g.,][]{p1}. 

The SHELLQs project has spectroscopically identified $\sim$180 broad-line quasars at $5.6 < z < 7.1$, in which 139 have been published so far
\citep{p1,p4,p2,p5,p10,p7,p16,p19,p20}.
In addition, we have discovered $\sim$100 luminous galaxies in the same redshift range, with 
no or only narrow (FWHM $<$ 500 km s$^{-1}$) Ly$\alpha$ emission.
Among these galaxies, 34 objects have very strong Ly$\alpha$, with a line luminosity $L_{\rm Ly\alpha} > 10^{43}$ erg s$^{-1}$.
Since such luminous Ly$\alpha$ emitters (LAEs) are often associated with AGN signatures at other wavelengths in the lower-$z$ universe \citep{konno16, sobral18, spinoso20}, we have suggested that they are candidate narrow-line quasars.
\citet{onoue21} detected strong \ion{C}{4} $\lambda$1549 doublet lines from one of them in deep near-IR spectroscopy with the Keck Telescope,
giving a strong indication for the presence of AGN.
The rest-UV spectra of the 34 galaxies are reminiscent of type-II quasar candidates at lower redshifts \citep{alexandroff13} selected from Sloan Digital Sky Survey \citep[SDSS;][]{york00}, 
although the latter objects have larger Ly$\alpha$ widths of FWHM $\sim$ 1000 km s$^{-1}$ \citep[see, e.g.,][]{p10}.

This paper describes rest-frame optical spectroscopy with JWST/NIRSpec for 13 targets from the SHELLQs sample, as summarized in Table \ref{tab:targets}. 
Among them, 11 objects (G01 -- G11) were drawn from the above sample of 34 galaxies with strong Ly$\alpha$.
One of them (G11) was targeted as part of the JWST General Observers (GO) program 1967 (PI: M. Onoue).
The other 10 targets (G01 -- G10) have the highest Ly$\alpha$ luminosity among the 34 galaxies, and were selected for dedicated follow-up observations with the JWST GO program 3417 (PI: Y. Matsuoka).
We also include two similar objects (Q01 and Q02) from GO 1967 in the present analysis.
Q02 has a narrow Ly$\alpha$ profile (see Table \ref{tab:targets}), 
but 
SHELLQs has classified it as a broad-line quasar since there is a signature of a broad component at the base of Ly$\alpha$ \citep[][see also \S \ref{sec:lya} of this paper]{p4}.
Q01 has a relatively broad Ly$\alpha$ (FWHM $\sim$1600 km s$^{-1}$), but its width is closer to the above targets than to 
most of the known EoR quasars. 
We see below that the rest-optical spectra of Q01 and Q02 look very similar to the majority of G01 -- G11 (see Figures \ref{fig:spec1} -- \ref{fig:spec3}),  
hence they may belong to the same parent population.


\subsection{Observations and Data Reduction}

The GO 1967 program was executed in Cycle 1.
NIRSpec was used with the medium-resolution grating G395M and transmission filter F290LP, which provide the spectral resolution $R \sim 1000$ over the
wavelength range from 2.87 to 5.10 $\mu$m. 
The S200A2 slit was used, and the total exposure time was 3000 -- 9000 sec per target.
More details of the observations and data reduction are given in M. Onoue et al. (in prep.), and also in \citet{ding23} and \citet{onoue24} for other targets from the same program.

The GO 3417 program was carried out in Cycle 2, during
a period from March to October 2024.
We used NIRSpec with the high-resolution grating G395H and transmission filter F290LP.
This configuration provides wavelength coverage from 2.87 to 5.14 $\mu$m (with a small gap from 3.81 to 3.92 $\mu$m for the slit we used), with the spectral resolution $R \sim 2700$.
The observations were carried out in the Fixed-Slits mode, with the S200A2 slit.
The slit is 0.\arcsec2 wide and 3.\arcsec3 long.
The total exposure time on source was 3700 sec per target, divided into 5 point dithers $\times$ 1 integration $\times$ 50 groups.
We exploited the NRSIRS2RAPID readout mode with the FULL array.
Target acquisition was performed with the wide aperture target acquisition (WATA) mode, using the F140X filter and NRSRAPIDD6 readout mode.  

The data from GO 3417 were reduced with the JWST pipeline version 1.13.4, using the reference files specified by \texttt{jwst\_1293.pmap}.
We kept the default parameter settings throughout the pipeline process.
A small number of pixels affected by ``snowball" features (caused by large cosmic ray impacts) and other apparent defects were manually spotted and excluded from the final stacking.

\begin{figure*}
\epsscale{1.1}
\plotone{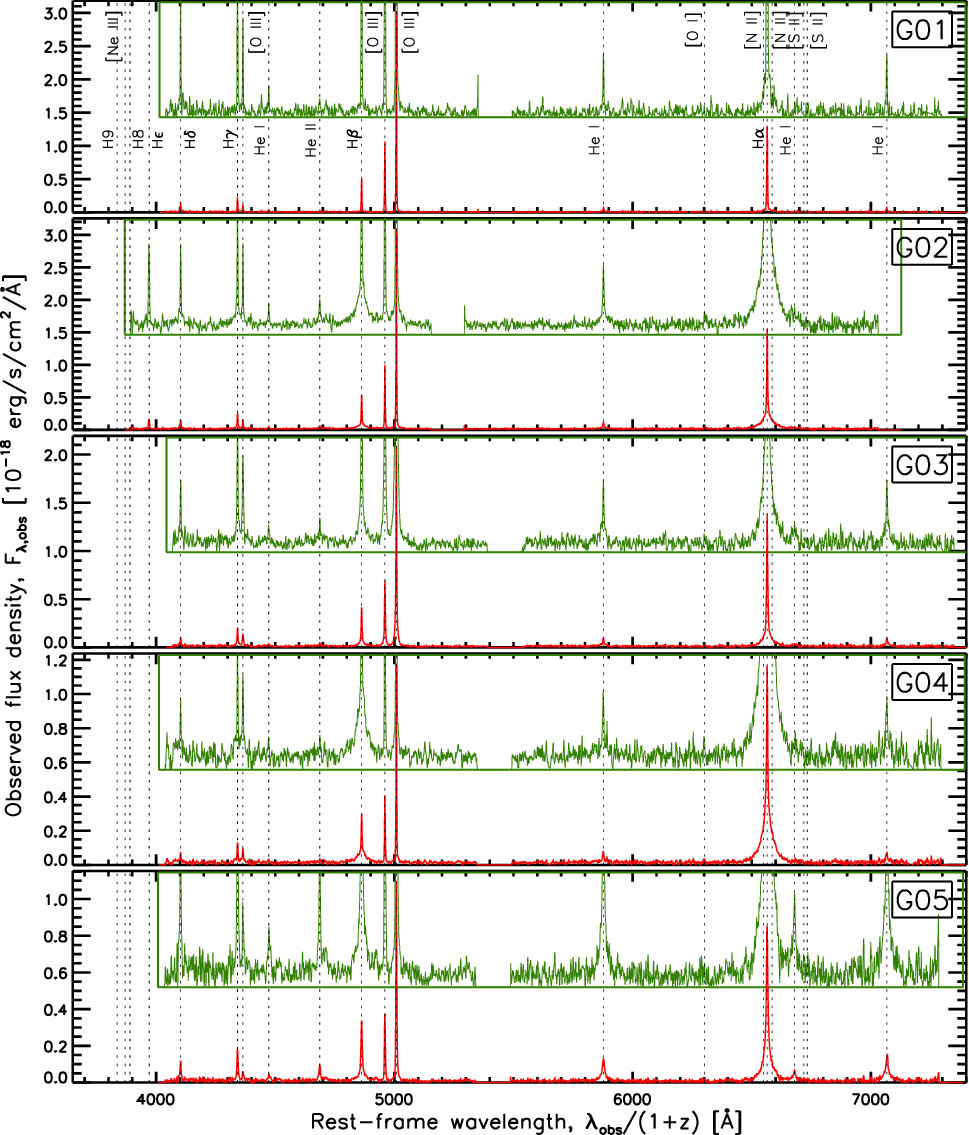}
\caption{
JWST/NIRSpec spectra of G01 -- G05 observed in GO 3417 (red lines). 
The flux density is given in the observer's frame, while the wavelength is given in the rest frame for the purpose of presentation, using the systemic redshifts determined from the spectral models (see \S \ref{sec:specmodel}).
The insets with the green lines show the same spectra plotted on an expanded scale, to make fainter emission lines apparent.
The dotted lines indicate the expected central wavelengths of common emission lines, as labeled in the top panel.
\label{fig:spec1}}
\end{figure*}

\begin{figure*}
\epsscale{1.1}
\plotone{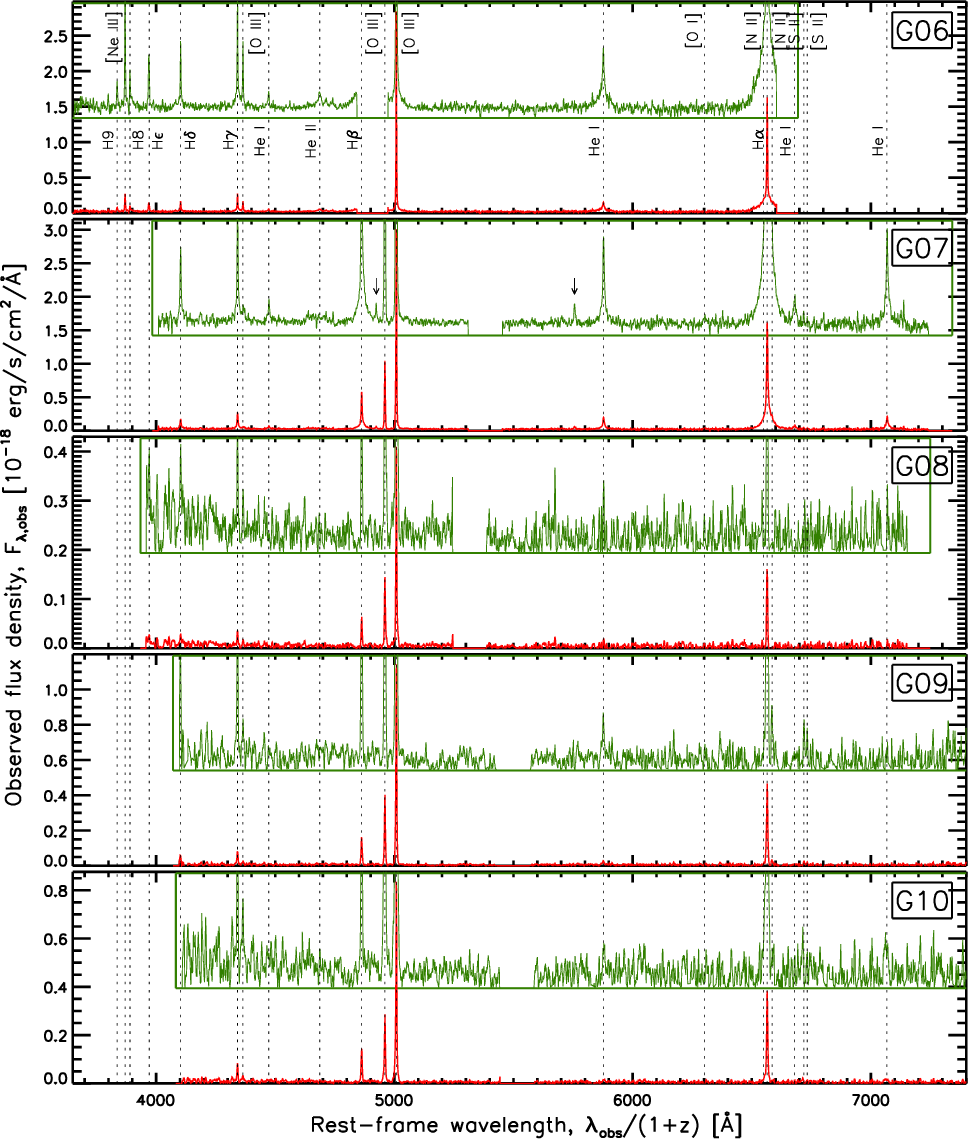}
\caption{
Same as Figure \ref{fig:spec1}, but for G06 -- G10 observed in GO 3417.
G07 has weak unidentified emission lines at 4924 \AA\ and 5756 \AA, as marked by the arrows.
The former wavelength is close to that of the \ion{Fe}{2} Opt 42 multiplet at 4931 \AA\ \citep{vandenberk01}.
The latter line may also be associated with the iron species, given that the two unidentified lines both appear in this object.
\label{fig:spec2}}
\end{figure*}

\begin{figure*}
\epsscale{1.1}
\plotone{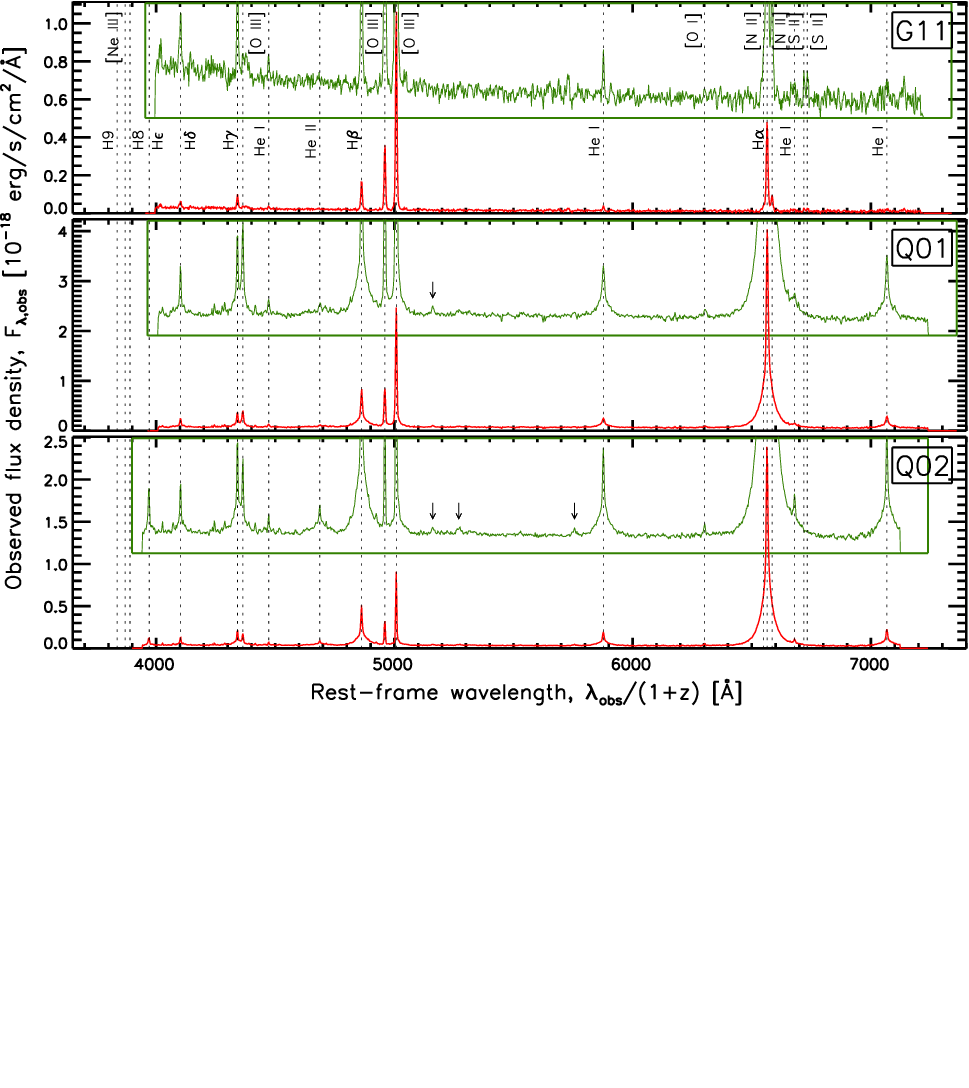}
\caption{
Same as Figure \ref{fig:spec1}, but for G11, Q01, and Q02  observed in GO 1967.
Q01 has an unidentified line at 5161 \AA\ (marked by the arrow), whose wavelength coincides with that of a [\ion{Fe}{7}] emission line \citep{vandenberk01}.
Q02 shares an unidentified line at 5161 \AA\ with Q01 and that at 5756 \AA\ with G07 (see Figure \ref{fig:spec2}).
An additional weak line is seen at 5270 \AA, which is close to the wavelength of [\ion{Fe}{7}] at 5278 \AA. 
\label{fig:spec3}}
\end{figure*}

Figures \ref{fig:spec1} -- \ref{fig:spec3} present the reduced spectra of the 13 targets, both from GO 1967 and GO 3417.
The \ion{H}{1} Balmer lines and [\ion{O}{3}] lines are very strong, while the continuum emission is relatively faint.
Weaker emission lines such as \ion{He}{1}, \ion{He}{2}, and [\ion{Ne}{3}] were also detected, but [\ion{N}{2}] and  [\ion{S}{2}] lines are absent in most objects.
It is also clear at first sight that many targets have a strong broad component in the \ion{H}{1} lines.

\bigskip

\subsection{Spectral Modeling} \label{sec:specmodel}

\subsubsection{Prescription}

In order to measure the properties of each emission component, we modeled and fitted the observed spectra.
The spectra were first converted to the rest frame by using the Ly$\alpha$ redshifts ($z_{\rm Ly\alpha}$) reported in Table \ref{tab:targets}.
The continuum emission was assumed to follow a power-law function, with the amplitude and slope varied as free parameters.
There may be significant contribution from starlight, but the limited wavelength coverage and signal-to-noise ratios (S/N) prevent us from assuming more complicated continuum models.

\begin{figure}
\epsscale{1}
\plotone{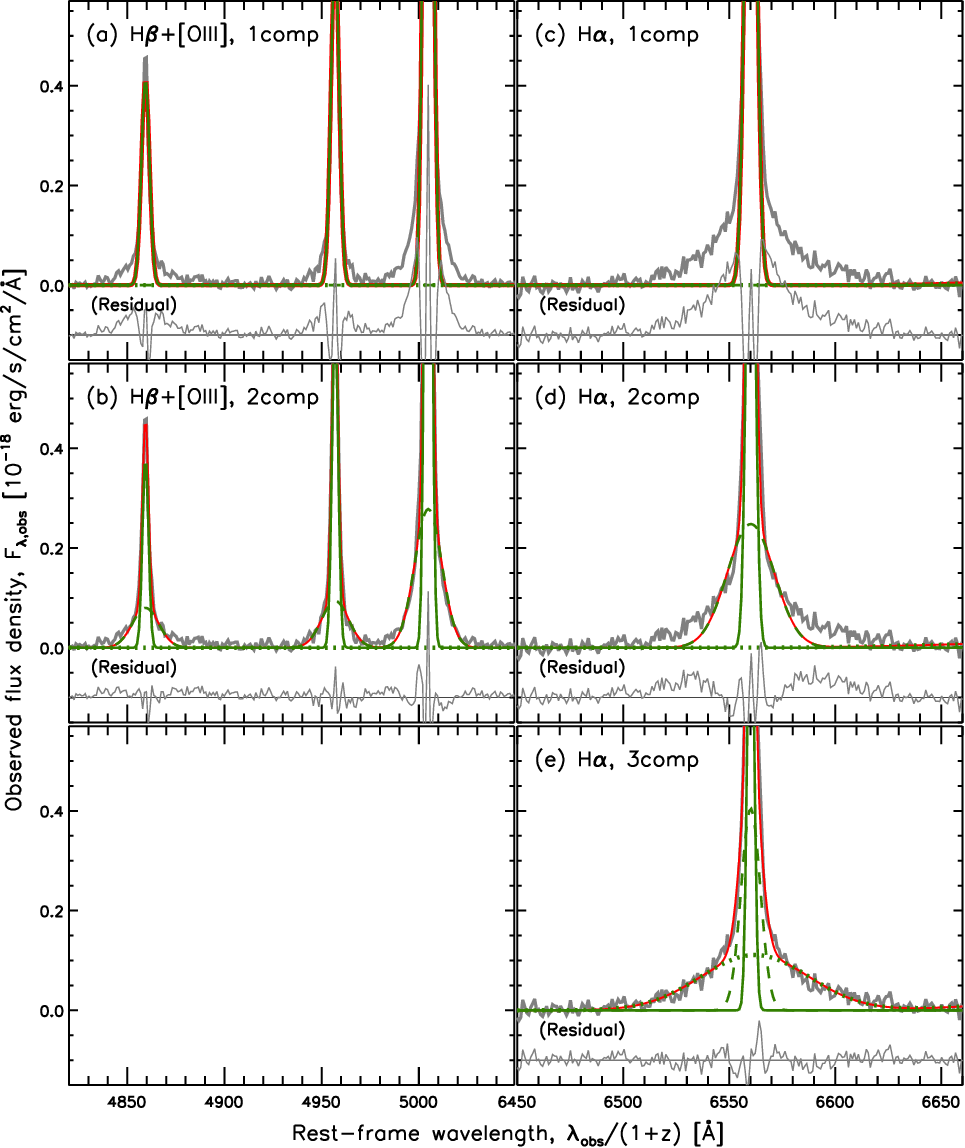}
\caption{
An example of spectral model fitting.
The thick gray lines represent the observed spectrum of G03 after subtracting the power-law continuum and \ion{Fe}{2} pseudo-continuum, 
around H$\beta$ $+$ [\ion{O}{3}] $\lambda$4959, $\lambda$5007 (left panels) and around H$\alpha$ (right panels).
The best-fit models (red lines) consist of one Gaussian component (top panels), two components (middle panels), or three components (bottom panel).
The green solid, dashed, and dotted lines represent the first, second, and third component, respectively.
The central wavelengths of the first and second components are tied together for a given line, while the third component is allowed to have a different central wavelength (see text).
The thin gray line in each panel represents the fitting residual, for which a vertical offset of $-0.1$ (the black horizontal line) was given to improve visibility. 
\label{fig:specfit_demo}}
\end{figure}

Each emission line was modeled as a sum of multiple Gaussian functions.
Figure \ref{fig:specfit_demo} presents a typical case of model fitting, demonstrating that at least two and three components are needed to reproduce the strongest forbidden ([\ion{O}{3}] $\lambda$5007; panels a and b) and permitted (H$\alpha$; panels c, d, e) lines, respectively.
We modeled the profile of forbidden lines ([\ion{Ne}{3}], [\ion{O}{3}], [\ion{O}{1}], [\ion{N}{2}], or [\ion{S}{2}]) with
\begin{eqnarray}
f_\lambda & = & F_{\rm 1}\ G \left(\lambda_0 \left(1 - \frac{v_{\rm 1}}{c}\right), w_{\rm 1}\right)\nonumber\\
                  & + & r_{\rm f}  F_{\rm 1}\ G \left(\lambda_0 \left(1 - \frac{v_{\rm 2}}{c}\right), w_{\rm 2}\right) ,
\label{eq:forbidden}
\end{eqnarray}
and the profile of permitted lines (\ion{H}{1}, \ion{He}{1}, or \ion{He}{2}) with
\begin{eqnarray}
f_\lambda & = & F_{\rm 1}\ G \left(\lambda_0 \left(1 - \frac{v_{\rm 1}}{c}\right), w_{\rm 1}\right)\nonumber\\
                  & + & r_{\rm p}  F_{\rm 1}\ G \left(\lambda_0 \left(1 - \frac{v_{\rm 2}}{c}\right), w_{\rm 2}\right)\nonumber\\
                  & + & F_{\rm 3}\ G \left(\lambda_0 \left(1 - \frac{v_{\rm 3}}{c}\right), w_{\rm 3}\right) .
\label{eq:permitted}
\end{eqnarray}
Here $G (l, w)$ represents a normalized Gaussian function with the central wavelength of $l$ (\AA) and FWHM of $w$ (km s$^{-1}$). 
The Gaussian amplitudes $F_{\rm 1}$ and $F_{\rm 3}$ were varied freely for the individual lines, while 
we assumed the fixed line ratios of [\ion{O}{3}] $\lambda$5007/$\lambda$4959 = 2.98 and [\ion{N}{2}] $\lambda$6585/$\lambda$6550 = 2.94 \citep[e.g.,][]{storey00, osterbrock06}.
Each of the line FWHMs ($w_{\rm 1}$, $w_{\rm 2}$, $w_{\rm 3}$; $w_{\rm 1} < w_{\rm 2} < w_{\rm 3}$), blueshift velocities relative to Ly$\alpha$ ($v_{\rm 1}$, $v_{\rm 2}$, $v_{\rm 3}$), and flux ratios ($r_{\rm f}$, $r_{\rm p}$) was tied as a single free parameter in a given object.
We applied the instrumental line broadening, using the disperser resolution as a function of wavelength taken from the JWST User Documentation (JDox\footnote{
https://jwst-docs.stsci.edu}).
The fixed values $\lambda_0$ and $c$ represent the intrinsic central wavelength of the line and the speed of light, respectively.

The spectral model also includes the \ion{Fe}{2} pseudo-continuum, assuming the \citet{boroson92} template with a varying amplitude.
Since \ion{Fe}{2} is thought to originate from the outermost part of AGN broad line region
\citep[BLR; see the related discussion below; e.g.,][]{matsuoka07,matsuoka08},
the template was velocity-broadened with a Gaussian kernel with the FWHM of $w_2$. 
It turned out that this emission component is very weak, 
and so it has little impact on the model fitting.
Indeed, we will see below that it has $>$5$\sigma$ significance of the integrated flux from $\lambda_{\rm rest}$ = 4434 to 4684 \AA\ \citep{boroson92} only in a few objects, 
and that it is weaker than a single H$\delta$ line in each of the objects.

\begin{deluxetable*}{cccccccc}
\tablecaption{Best-fit parameters of the spectral models \label{tab:velocities}}
\tablewidth{0pt}
\tablehead{
\colhead{Name} & \colhead{$z_{\rm sys}$} & \colhead{$r_{\rm f}$} & \colhead{$r_{\rm p}$} &   \colhead{$w_{\rm 1}$} & \colhead{$w_{\rm 2}$} & \colhead{$w_{\rm 3}$} & \colhead{$v_{\rm 3} - v_{\rm 1}$}\\
\colhead{} & \colhead{} & \colhead{} & \colhead{} & \colhead{(km s$^{-1}$)} & \colhead{(km s$^{-1}$)} & \colhead{(km s$^{-1}$)} & \colhead{(km s$^{-1}$)} 
}
\startdata
G01 &  6.127 &  0.08 $\pm$ 0.01  & 0.12 $\pm$ 0.02 &    145 $\pm$    1 &    759 $\pm$   35 &  2440 $\pm$  190 &   $-$98 $\pm$   1\\
G02 &  6.397 &  0.32 $\pm$ 0.01  & 0.92 $\pm$ 0.03 &    113 $\pm$    1 &    470  $\pm$   7  &  3245 $\pm$   36  &   $-$66 $\pm$  80\\
G03 &  6.075 &  0.64 $\pm$ 0.01  & 0.79 $\pm$ 0.02 &    184 $\pm$    1 &    944 $\pm$   10 &  3650 $\pm$  110 & $-$160 $\pm$ 120\\
G04 &  6.130 &  0.18 $\pm$ 0.01  & 1.60 $\pm$ 0.05 &    153 $\pm$    2 &    855 $\pm$   19 &  3496 $\pm$   46  &    22  $\pm$  1\\
G05 &  6.139 &  0.26 $\pm$ 0.01  & 1.55 $\pm$ 0.06 &    227 $\pm$    2 &    783  $\pm$  18 &  3120 $\pm$   66 &    $-$37  $\pm$  1\\
G06  & 6.875 &  0.20 $\pm$ 0.01  & 0.80 $\pm$ 0.03 &    136 $\pm$    1 &  1090  $\pm$  23 &  4191 $\pm$   95  &   $-$24  $\pm$ 25\\
G07 &  6.181 &  0.33 $\pm$ 0.01  & 1.22 $\pm$ 0.04 &    174 $\pm$    2 &   520  $\pm$   8   &  2396 $\pm$   24 &  $-$164  $\pm$  9\\
G08 &  6.273 &  0.44 $\pm$ 0.04  & 0.21 $\pm$ 0.07 &    143 $\pm$    4 &   553  $\pm$  31  &   \nodata & \nodata\\
G09 &  6.027 &  0.67 $\pm$ 0.07  & 0.77 $\pm$ 0.10 &    177 $\pm$    5 &   411  $\pm$  12 &   \nodata & \nodata\\
G10 &  6.007 &  0.62 $\pm$ 0.02  & 0.69 $\pm$ 0.04 &    137 $\pm$    3 &   570  $\pm$  12 &   \nodata & \nodata\\
G11  & 6.203 &  0.22 $\pm$ 0.02  & 0.42 $\pm$ 0.03 &    203 $\pm$    6 &   862  $\pm$  35 &   \nodata & \nodata\\
Q01 &  6.183 &  0.39 $\pm$ 0.01  & 1.66 $\pm$ 0.02 &    267 $\pm$    2 &  2005  $\pm$  14  & 5333 $\pm$   39 &  $-$134 $\pm$   9\\
Q02 &  6.302 &  0.31 $\pm$ 0.01  & 2.00 $\pm$ 0.01 &    163 $\pm$    2 &  1685  $\pm$  10  & 4872 $\pm$   18 &   $-$70  $\pm$  7\\
\enddata
\tablecomments{
The line FWHMs ($w_1$, $w_2$, $w_3$) have been corrected for the instrumental resolution.
The last column lists the blueshift velocities of the broadest Gaussian component, relative to the narrower components (which define the systemic redshifts).}
\end{deluxetable*}

\subsubsection{Fitting to the Data}

The model fitting was performed with the least-$\chi^2$ method, using the pixel-to-pixel errors taken from the JWST pipeline output.
All the free parameters were varied simultaneously.
We include the third Gaussian component of permitted lines only when it has more than 5$\sigma$ significance and the difference in 
the Bayesian information criteria \citep[BIC; e.g.,][]{liddle07}, defined as 
$\Delta$BIC $\equiv$ BIC$_{\rm 2G}$ $-$ BIC$_{\rm 3G}$ (where BIC$_{\rm 2G}$ and BIC$_{\rm 3G}$ represent the BIC for models without and with the third Gaussian component 
in Equation \ref{eq:permitted}, respectively) exceeds 10.
A reasonable fit was obtained when $v_1$ and $v_2$ are tied together, 
so we set $v_1 = v_2$ in every object in order to reduce the number of free parameters and avoid overfitting.
The parameter $v_3$ was not tied to $v_1$ or $v_2$.
We define the systemic redshift ($z_{\rm sys}$) with the central wavelengths of the narrower Gaussian components, using the best-fit values of $v_1 = v_2$.
Table \ref{tab:velocities} lists the determined values, as well as the best-fit parameters describing the overall line profiles of the individual objects.

\begin{figure*}
\epsscale{1.0}
\plotone{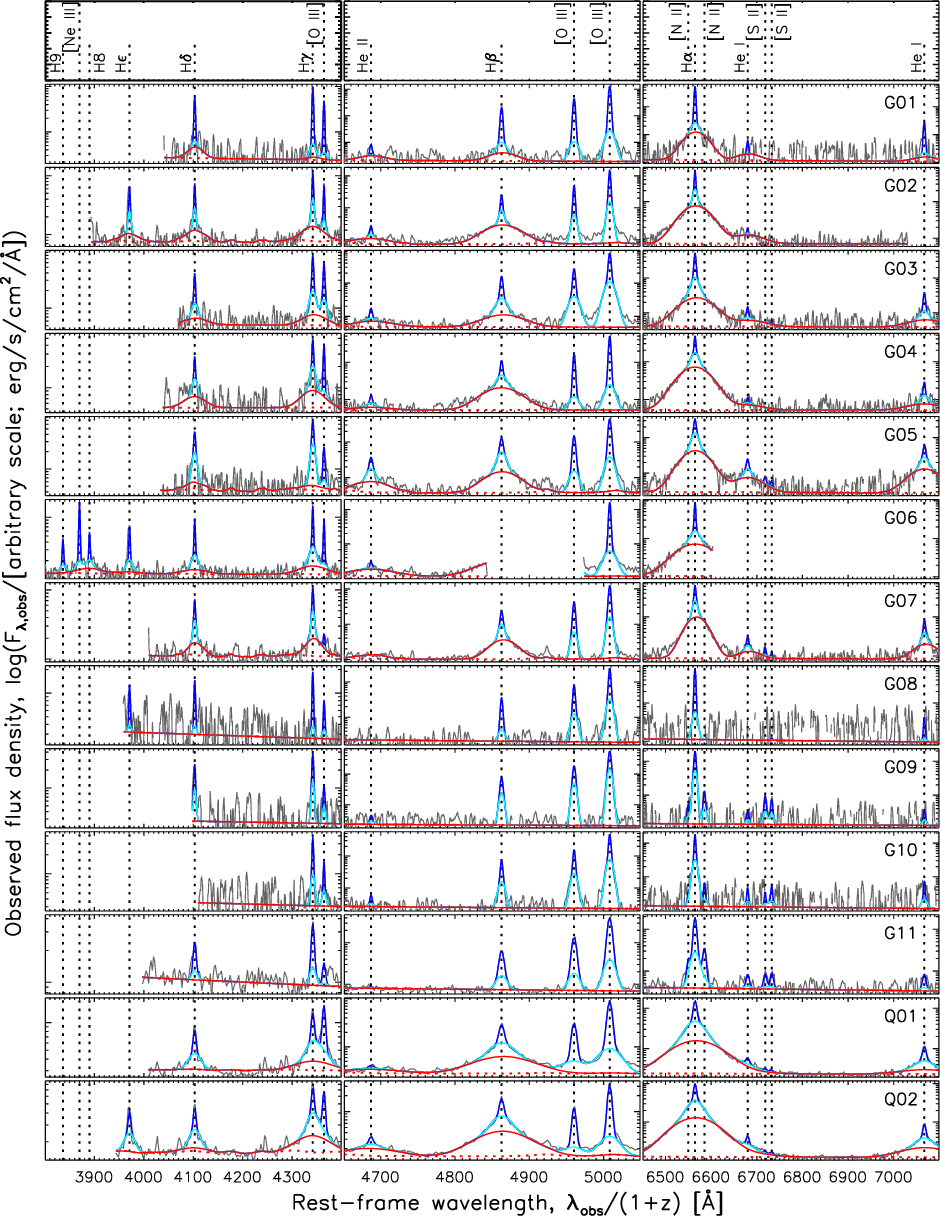}
\caption{
Best-fit spectral models, composed of the power-law continuum ($f_{\rm pl}$), \ion{Fe}{2} pseudo-continuum ($f_{\rm FeII}$), and three 
Gaussian components ($f_{\rm G1}$, $f_{\rm G2}$, and $f_{\rm G3}$, in order of increasing line width).
Note that the fluxes are plotted on a logarithmic scale.
The gray lines represent the observed spectra.
The red dotted lines represent $f_{\rm pl} + f_{\rm FeII}$, while the red, cyan, and blue solid lines represent 
$f_{\rm pl} + f_{\rm FeII} + f_{\rm G1}$, $f_{\rm pl} + f_{\rm FeII} + f_{\rm G1} + f_{\rm G2}$, and 
$f_{\rm pl} + f_{\rm FeII} + f_{\rm G1} + f_{\rm G2} + f_{\rm G3}$ (total model), respectively.
The cyan/red lines are drawn on top of the blue/cyan lines, for better visibility of the individual line components.
The vertical dotted lines indicate the expected central wavelengths of emission lines, as labeled in the top panel.
\label{fig:specfit}}
\end{figure*}

The best-fit spectral models are presented in Figure \ref{fig:specfit}.
In all cases the two narrower Gaussian components are tightly constrained by the bright [\ion{O}{3}] $\lambda$ 5007 line, which has very high S/N.
For G08, G09, G10, and G11, all the emission lines can be adequately reproduced with these two components.\footnote{
G08, G09, and G10 have $\Delta$BIC $<$ 0. G11 has $\Delta$BIC = 10 and 4.5$\sigma$ significance of the third Gaussian component, and thus is a rather ambiguous case.
It could have a broad component depending on assumptions and model prescriptions (e.g., M. Onoue et al., in prep.), but here we conservatively treat it as objects with only narrow lines.
The classification of this single object doesn't affect the conclusions of this paper.
}
We found $r_{\rm f} \simeq r_{\rm p}$ in these cases (see Table \ref{tab:velocities}), which suggests that the forbidden and permitted lines have similar profiles.
The two components have FWHMs of $w_1$ = 100 -- 200 km s$^{-1}$ and $w_2$ = 400 -- 900 km s$^{-1}$, respectively.\footnote{
The NIRSpec disperser resolution is $\sim$100 km s$^{-1}$ for G395H and $\sim$300 km s$^{-1}$ for G395M, so these narrowest components are not fully resolved. See also a related discussion later in this section.}
We interpret that all these lines originate from star-forming (SF) regions of the galaxies or from narrow line region (NLR) associated with AGNs.
For simplicity, the four objects are called ``narrow-line (NL) objects" hereafter.
On the other hand, it is quite obvious that the permitted lines of the remaining nine objects (G01 -- G07, Q01, and Q02) require an additional broader component. 
The FWHMs of this third component are $w_3$ = 2000 -- 5000 km s$^{-1}$.
As we will discuss in \S \ref{sec:nature}, these objects are most likely contributed by emission from BLR associated with AGNs. 
We call them ``broad-line (BL) objects" hereafter.
The BL objects belong to the population of BHEs, but they have broad components in other Balmer lines and \ion{He}{1} lines in addition to H$\alpha$ and H$\beta$.

The disperser resolution provided in the JDox has been estimated for a fully illuminated slit, so may be different from what should be applied for point sources. 
\citet{degraaff24} modeled the point source resolution of the G395H disperser, and reported that it can be higher than the JDox values by nearly a factor of two.
We conducted a test run of the spectral modeling with their resolution model, finding that $w_1$ increases by $\sim$20 \% ($w_2$ and $w_3$ increase by smaller amounts).
Such a modification changes none of our conclusions presented below. 

\begin{deluxetable*}{ccccccc}
\tablecaption{Emission line properties and dust extinction\label{tab:ew_av}}
\tablewidth{0pt}
\tablehead{
\colhead{Name} & \colhead{EW$_{\rm [O III]}$} &\colhead{EW$_{\rm H\alpha, n}$} & \colhead{EW$_{\rm H\alpha, b}$} & \colhead{FWHM$_{\rm H\alpha, b}$} &   \colhead{$A_{V, \rm {n}}$} & \colhead{$A_{V, \rm{b}}$} \\
\colhead{} & \colhead{(\AA)} & \colhead{(\AA)} & \colhead{(\AA)} & \colhead{(km s$^{-1}$)} & \colhead{} & \colhead{} 
}
\startdata
G01  &     2520 $\pm$ 850 &1800 $\pm$ 460 &  580 $\pm$ 160  & 1855 $\pm$   94 &  0.10 $\pm$ 0.04  &  2.15 $\pm$ 0.56\\
G02  &      745  $\pm$ 12   &  313 $\pm$   7   &  878 $\pm$  16   &  2370 $\pm$   21 &  0.46 $\pm$ 0.06 &   1.01 $\pm$ 0.06\\
G03  &      958  $\pm$ 44   &  812 $\pm$  30  &  557 $\pm$  27   &  2542 $\pm$   40 &  0.94 $\pm$ 0.05 &   1.40 $\pm$ 0.17\\
G04  &      374  $\pm$   6   &  278 $\pm$   6   & 1356 $\pm$  26  &  2388 $\pm$   17 &  1.75 $\pm$ 0.08 &   1.88 $\pm$ 0.07\\
G05  &      654  $\pm$ 59   &  467 $\pm$  35  & 1460 $\pm$ 110 &  1810 $\pm$   19 &  0.28 $\pm$ 0.10 &   1.03 $\pm$ 0.11\\
G06  &      518  $\pm$   9   &  300 $\pm$   5   &   716 $\pm$  17  &  2840 $\pm$   35 &  \nodata               &   1.52 $\pm$ 0.12\\
G07  &      570  $\pm$ 15   &  257 $\pm$   7   &   748 $\pm$  18  &  1736 $\pm$   12 &  0.48 $\pm$ 0.08 &   1.13 $\pm$ 0.07\\
G08  &      520  $\pm$ 190 &  319 $\pm$  93  &  \nodata              &   \nodata               &  0.00 $\pm$ 0.19 &   \nodata\\
G09  &     1300 $\pm$ 300 &  830 $\pm$ 160 &  \nodata              &   \nodata               &  0.37 $\pm$ 0.21 &   \nodata\\
G10  &      920  $\pm$ 290 &  750 $\pm$ 180 &  \nodata              &   \nodata               & 0.13 $\pm$ 0.10  &   \nodata\\
G11  &      435  $\pm$ 29   &  435 $\pm$  25  &  \nodata              &   \nodata               &  0.66 $\pm$ 0.09 &   \nodata\\
Q01  &      298  $\pm$  3    &  523 $\pm$   5   & 1161 $\pm$  12  &   3600 $\pm$    8 &  2.56 $\pm$ 0.03  &  1.94 $\pm$ 0.04\\
Q02  &      176  $\pm$  2    &  475 $\pm$   4   & 1682 $\pm$  11  &   3415 $\pm$    6 &  2.62 $\pm$ 0.03  &  2.09 $\pm$ 0.02\\
\enddata
\tablecomments{
The EWs of [\ion{O}{3}] $\lambda$5007 (EW$_{\rm [O III]}$) and H$\alpha$ (EW$_{\rm H\alpha, n}$ and EW$_{\rm H\alpha, b}$) are reported in the rest frame.
The FWHMs of the H$\alpha$ BLR component (FWHM$_{\rm H\alpha, b}$) have been corrected for the instrumental resolution.}
\end{deluxetable*}

\begin{deluxetable*}{cccccccc}
\tablecaption{Luminosities and black hole masses\label{tab:deliverables}}
\tablewidth{0pt}
\tablehead{
\colhead{Name} & \colhead{$\log L_{5100}$} & \colhead{$\log L_{\rm [OIII]}$} & \colhead{$\log L_{\rm H\alpha, n}$} & \colhead{$\log L_{\rm H\alpha, b}$}
& \colhead{$\log L_{\rm bol}$} & \colhead{$\log M_{\rm BH}$} & \colhead{$\lambda_{\rm Edd}$}
}
\startdata
G01 &  41.13 $\pm$ 0.25  & 43.64 $\pm$ 0.02 &  43.29 $\pm$ 0.01 &  43.43 $\pm$ 0.18 &  45.55 $\pm$ 0.18 &   7.80 $\pm$ 0.10 &   0.45 $\pm$ 0.21\\   
G02 &  41.23 $\pm$ 0.03  & 43.86 $\pm$ 0.03 &  43.42 $\pm$ 0.02 &  44.04 $\pm$ 0.02 &  46.15 $\pm$ 0.02 &   8.30 $\pm$ 0.01 &   0.56 $\pm$ 0.03\\   
G03 &  41.20 $\pm$ 0.08  & 43.99 $\pm$ 0.02 &  43.72 $\pm$ 0.02 &  43.70 $\pm$ 0.06 &  45.82 $\pm$ 0.06 &   8.21 $\pm$ 0.03 &   0.33 $\pm$ 0.05\\   
G04 &  41.47 $\pm$ 0.03  & 44.00 $\pm$ 0.04 &  43.63 $\pm$ 0.03 &  44.36 $\pm$ 0.02 &  46.47 $\pm$ 0.02 &   8.46 $\pm$ 0.01 &   0.82 $\pm$ 0.05\\   
G05 &  40.92 $\pm$ 0.05  & 43.41 $\pm$ 0.05 &  43.14 $\pm$ 0.03 &  43.87 $\pm$ 0.04 &  45.98 $\pm$ 0.04 &   7.98 $\pm$ 0.02 &   0.80 $\pm$ 0.08\\   
G06 &  41.67 $\pm$ 0.05  & 44.40 $\pm$ 0.06 &  43.91 $\pm$ 0.04 &  44.29 $\pm$ 0.04 &  46.41 $\pm$ 0.04 &   8.58 $\pm$ 0.02 &   0.53 $\pm$ 0.05\\   
G07 &  41.41 $\pm$ 0.03  & 43.88 $\pm$ 0.04 &  43.42 $\pm$ 0.03 &  44.08 $\pm$ 0.02 &  46.20 $\pm$ 0.02  &  8.04 $\pm$ 0.01 &   1.13 $\pm$ 0.07\\   
G08 &  40.20 $\pm$ 0.09  & 42.93 $\pm$ 0.09 &  42.47 $\pm$ 0.07 &   \nodata                 &  \nodata   & \nodata       &  \nodata\\  
G09 &  40.38 $\pm$ 0.09  & 43.51 $\pm$ 0.10 &  43.10 $\pm$ 0.07 &   \nodata                 &  \nodata   & \nodata       &  \nodata\\  
G10 &  40.31 $\pm$ 0.04  & 43.29 $\pm$ 0.05 &  42.94 $\pm$ 0.03 &   \nodata                 &  \nodata   & \nodata       &  \nodata\\  
G11 &  41.04 $\pm$ 0.04  & 43.70 $\pm$ 0.04  & 43.36 $\pm$ 0.03  &  \nodata                 &  \nodata   & \nodata       &  \nodata\\  
Q01 &  42.21 $\pm$ 0.02  & 44.99 $\pm$ 0.01 &  44.81 $\pm$ 0.01 &  44.97 $\pm$ 0.01 &  47.08 $\pm$ 0.01 &   9.11 $\pm$ 0.01 &   0.74 $\pm$ 0.02\\   
Q02 &  41.98 $\pm$ 0.01  & 44.50 $\pm$ 0.01 &  44.51 $\pm$ 0.01 &  44.89 $\pm$ 0.01 &  47.01 $\pm$ 0.01 &   9.03 $\pm$ 0.01 &   0.76 $\pm$ 0.01\\   
\enddata
\tablecomments{
The luminosities and black hole masses are presented in units of erg s$^{-1}$ and $M_\odot$, respectively.
All the quantities in this table have been corrected for dust extinction.
The reported errors do not include uncertainties inherent in the assumed bolometric correction and black hole mass calibration (see \S \ref{sec:luminosities}).
}
\end{deluxetable*}


Tables \ref{tab:line_fluxes1} and \ref{tab:line_fluxes2} in Appendix \ref{sec:appendix} report the measured line fluxes and continuum properties. 
We attribute all the emission lines of the NL objects, and all the forbidden lines of the BL objects, to emission from SF/NLR.
The permitted lines of the BL objects are decomposed into the contributions from SF/NLR and BLR emission, assuming that the former shares the same profile with the forbidden lines.
In practice, this was performed by attributing the flux of $f_{\rm G1} + r_{\rm f} f_{\rm G2}$ to SF/NLR emission and the flux of $(r_{\rm p} - r_{\rm f}) f_{\rm G2} + f_{\rm G3}$ to BLR emission,
where $f_{\rm G1}$, $f_{\rm G2}$, and $f_{\rm G3}$ represent the three Gaussian components in order of increasing width.
Hereafter, we refer to the quantities derived for the SF/NLR and BLR emission with subscripts of ``n" and ``b" (e.g., $A_{V, {\rm n}}$ and $A_{V, {\rm b}}$ for dust extinction), respectively.
The resultant rest-frame equivalent widths (EWs) of [\ion{O}{3}] $\lambda$5007 and H$\alpha$ are listed in Table \ref{tab:ew_av}.
Finally, we measured the FWHM of the BLR H$\alpha$ emission (FWHM$_{\rm H\alpha, b}$) in each BL object, by re-fitting a single Gaussian function to the decomposed H$\alpha$ profile (i.e., $(r_{\rm p} - r_{\rm f}) f_{\rm G2} + f_{\rm G3}$).

\subsubsection{Residual flux around [\ion{O}{3}] $\lambda$4959, $\lambda$5007}

While our spectral models can reproduce the observed spectra overall, we found non-negligible residual flux around [\ion{O}{3}] $\lambda$4959, $\lambda$5007.
This residual component appears only in several BL objects, such as G02, G03, G06, Q01, and Q02 (see Figure \ref{fig:specfit}). 
Therefore the presence of AGN may be related to its origin, e.g., AGN-driven [\ion{O}{3}] outflows or non-modeled \ion{Fe}{2} emission.
If we add a third Gaussian component to account for the residual in each of the two [\ion{O}{3}] lines (with the fixed flux ratio of $\lambda$5007/$\lambda$4959 = 2.98) in the model fitting, then the inferred line FWHMs are in the range of 1500 -- 6100 km s$^{-1}$.
It would indicate that fast outflows of highly-ionized, metal-enriched gas are present in the NLR.
If we further assume that H$\alpha$ line has the same outflowing NLR component (sharing the same flux ratio to the rest of the NLR emission with the [\ion{O}{3}] lines), then the estimated FWHM$_{\rm H\alpha, b}$ is reduced by up to $\sim$20 \%.
However, the impact of having this additional Gaussian is limited, and it doesn't change the need for broad line component in H$\alpha$ for the individual BL objects.


\section{Results and Discussion} \label{sec:results}

\subsection{Physical nature of the 13 objects} \label{sec:nature}

JWST has serendipitously discovered a number of BHEs among galaxies in the EoR, as we described in  \S \ref{sec:intro}. 
Our nine BL objects show similar properties to these BHEs, i.e., a broad component with FWHM $\simeq$ 2000 -- 5000 km s$^{-1}$ that are present only in rest-frame optical permitted lines. 
The high S/N of our spectra allows us to identify broad components from \ion{He}{1} lines, in addition to the \ion{H}{1} Balmer lines. 
The origin of such broad emission has been discussed extensively in the literature.
Starburst galaxies sometimes exhibit broad lines due to gas outflows, but their FWHMs are less than 1000 km s$^{-1}$ \citep[e.g.,][]{genzel11,newman12,swinbank19}.
Extreme cases with outflows exceeding 1000 km s$^{-1}$ are usually thought to be a signature of an AGN  \citep[e.g.,][]{shapiro09, zakamska16}.
Moreover, galactic outflows are typically observed both in forbidden and permitted lines \citep[e.g.,][]{freeman19}.
The associated signatures are prominent particularly in the [\ion{O}{3}] $\lambda$5007 line, due to enhanced ionization state and metal enrichment expected in the outflowing gas.
The spectra of our BL objects do not align with the picture of starburst-driven outflows described above, and thus our interpretation is that the broad emission has a non-stellar origin, for which the most likely explanation is BLR emission from an AGN.

The selection conditions of our BL objects are much different from those of other BHEs.
Most of the BHEs in the literature were found from blank-field surveys in JWST, over a few 100 arcmin$^2$ at most \citep[but see also][]{ma25b}. 
On the other hand, our targets have been pre-selected from a known sample of EoR galaxies with the highest Ly$\alpha$ luminosities, established by the HSC-SSP survey over 
$>$1000 deg$^2$. 
Hence our targets are by far the most luminous among EoR galaxies observed by JWST, as seen in Figure \ref{fig:agn_rel}.

\begin{figure*}
\epsscale{1.1}
\plotone{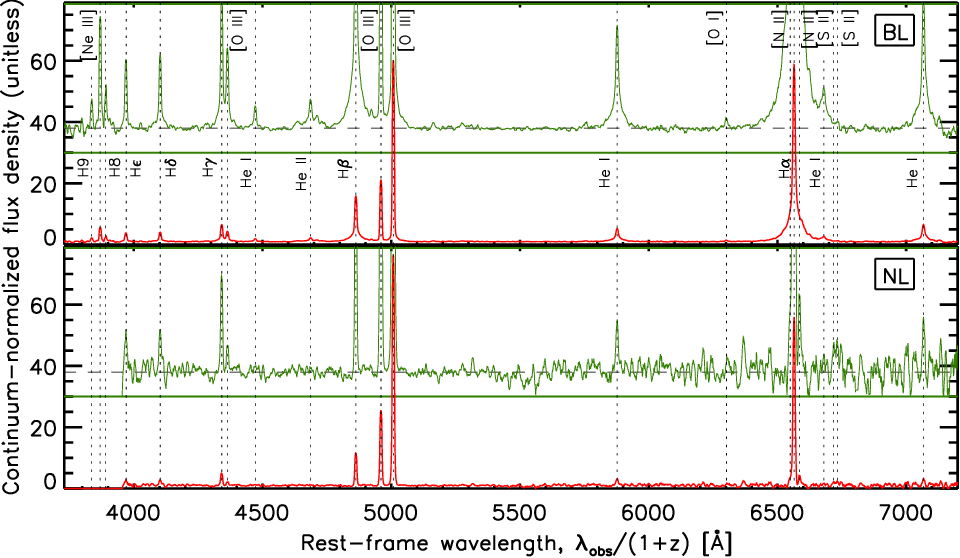}
\caption{
Stacked continuum-normalized spectra of the BL (top panel) and NL (bottom panel) objects. 
The insets with the green lines show the same spectra plotted on an expanded scale (an identical scaling was used for the BL and NL spectra), to make fainter emission lines apparent.
The dashed horizontal lines in the insets represent the normalized flux density of unity, i.e., the power-law continuum model (see text).
The spectra in these panels were smoothed with a boxcar kernel of 9 pixels, for display purposes.
The dotted lines indicate the expected central wavelengths of emission lines, as labeled in the top panel.
\label{fig:stack}}
\end{figure*}

The remaining four objects in our sample (NL objects) do not have broad line components, but they have extremely high [\ion{O}{3}]/H$\beta$ ratios ranging from 6.4 to 7.9.
If we use the BPT diagram \citep{baldwin81} with the classical \citet{kewley01} or \citet{kauffmann03} criteria, we could almost immediately rule out star-forming galaxies (SFGs) as their excitation sources. 
However, JWST has revealed that the SFG/AGN classification on the BPT diagram is much more complicated in the EoR,
due to low metallicity and high ionization state of the interstellar medium \citep[e.g.,][]{kocevski23, ubler23}.
Attempts have been made to overcome this issue, using new classification criteria with the line ratios of, e.g., 
[\ion{O}{3}] $\lambda$5007/H$\beta$, [\ion{N}{2}]/H$\alpha$,  [\ion{S}{2}]/H$\alpha$, [\ion{O}{1}]/H$\alpha$, [\ion{O}{3}] $\lambda$4363/H$\gamma$, and 
 [\ion{O}{3}] $\lambda$5007/4363 \citep{scholtz23,mazzolari24b,mazzolari24c}, aided by
photo-ionization models \citep[e.g.,][]{feltre16, gutkin16, nakajima22}.
We found that the four NL objects lie close to the borderlines between SFGs and AGNs on these diagnostic diagrams. 
Since many other BHEs are found within the SFG region with these 
classifications \citep[e.g.,][]{harikane23, maiolino24}, the dominant energy source of our NL objects remains unclear at the moment.

In Figure \ref{fig:stack}, we show the stacked continuum-normalized spectra of the BL and NL objects. 
These spectra were generated by dividing the individual spectra by the power-law continuum model, and then median-stacking 
in the rest frame.
The figure demonstrates higher EWs of most of the emission lines in the BL objects than in the NL objects, while [\ion{N}{2}] and [\ion{S}{2}] lines appear to be more prominent in the NL objects. 
A small spectral bump is seen in the BL objects at the base of \ion{He}{2} $\lambda$4687, in addition to that at the base of [\ion{O}{3}] $\lambda$ 4959 and $\lambda$5007, whose origins are unclear.
A major contribution from \ion{Fe}{2} is unlikely, since it would be accompanied by bands of strong emission at 4500 -- 4700 \AA\ and 5100 -- 5400 \AA\
\citep[e.g.,][]{boroson92, veron04, tsuzuki06}.
Further detailed analysis of the line fluxes and ratios, with dedicated photo-ionization model calculations, will be presented in papers in preparation.

\subsection{Dust extinction and luminosity} \label{sec:luminosities}

Tables \ref{tab:ew_av} and \ref{tab:deliverables} list the basic properties of the 13 objects, derived from the best-fit spectral models.
Dust extinction was estimated with the Balmer decrement separately for SF/NLR and BLR, assuming the intrinsic flux ratio of $F_{\rm H\alpha}$/$F_{\rm H\beta}$ = 2.86 and 
the extinction law of the Small Magellanic Cloud \citep[SMC;][]{pei92}.
While the listed values are the best possible estimates from the existing data, we note that the intrinsic Balmer decrement can deviate from the above value
for a variety of reasons, such as self-absorption of the line photons in dense BLR gas \citep[e.g.,][]{korista04}.
We detected significant amount of dust extinction with $A_{V, \rm{b}}$ $\sim$ 1 -- 2 mag, which overlaps with the distribution of the values measured for
BHEs/LRDs in the literature \citep[e.g.,][]{casey24, greene24}.
The extinction is larger toward BLR than toward SF/NLR in most cases, which is reasonable considering that the former is embedded deeper in the galaxies.
The Ly$\alpha$ emission from BLR would be subject to severe extinction, with $A_{\rm Ly\alpha}$ ranging from 6 to 13 mag (assuming the SMC extinction curve), and thus is practically obscured from the observer's line of sight. 
This is likely the reason that the BL objects G01 -- G07 were classified as galaxies, based on rest-UV spectroscopy with which they were discovered.
The SF/NLR extinction is also significant in many objects, although we see luminous narrow Ly$\alpha$ in the rest-UV spectra.
These Ly$\alpha$ emissions
may originate from the outskirts of the SF/NLR with smaller dust extinction, or from part of the entire SF/NLR with patchy dust distribution.

Table \ref{tab:deliverables} summarizes the luminosity measurements.
The continuum luminosity at 5100 \AA\ ($L_{5100}$, measured from the power-law continuum model) and the BLR H$\alpha$ luminosity ($L_{\rm H\alpha, b}$) 
have been corrected for dust extinction with $A_{V, \rm{b}}$ when it is available, otherwise with $A_{V, \rm{n}}$.
The [\ion{O}{3}] $\lambda$5007 and NLR H$\alpha$ luminosities ($L_{\rm [O III]}$ and $L_{\rm H\alpha, b}$) have been corrected with $A_{V, \rm{n}}$ 
($A_{V, \rm{b}}$ for G06).
The bolometric luminosity ($L_{\rm bol}$) of the BL objects was derived from the BLR H$\alpha$ luminosity, with the bolometric correction taken from \citet{stern12}.
The estimated values amount to $L_{\rm bol} = 10^{45.6 - 47.1}$ erg s$^{-1}$, which are comparable to those of 
SDSS quasars at lower redshifts \citep[e.g.,][]{shen11, rakshit20}. 
They are also around the break of the bolometric quasar luminosity function measured at $z = 6$ \citep[e.g.,][]{shen20}.
Therefore, we conclude that the present sample (at least the BL objects) represents the population of long-sought UV-obscured counterparts to luminous quasars in the EoR.

The NL objects (G08 -- G11) have lower line luminosities than do the BL objects, indicating a lower level of SMBH mass accretion, if indeed they are AGNs.
Alternatively, if the emission lines originate from SF regions, then the star formation rates inferred from $L_{\rm H\alpha, n}$ are 20 -- 200 $M_\odot$ yr$^{-1}$
with the \citet{kennicutt98} calibration.
If these objects belong to the main sequence of SFGs at $z \sim 6$ \citep[e.g.,][]{clarke24}, their stellar masses are estimated to be $>10^{10} M_\odot$.
These values are comparable to the host stellar masses of broad-line quasars at similar redshifts, measured from GO 1967 \citep{ding23,ding25}.

It is interesting to note that Q01 and Q02 have the largest dust extinction, and thus Ly$\alpha$ is expected to be most obscured, while they are the only two quasars (as identified from their Ly$\alpha$ emission)
in the present sample.
From the observed peak flux density of the BLR component of H$\alpha$, 
we estimate that of Ly$\alpha$ to be
$\sim$2 $\times$ 10$^{-22}$ erg s$^{-1}$ cm$^{-2}$ \AA$^{-1}$, assuming the intrinsic $F_{\rm H\alpha}$/$F_{\rm Ly\alpha}$ ratio of 0.3 \citep{vandenberk01} and accounting for dust extinction estimated above.
This flux density is far smaller than the errors of our discovery spectra. 
The broad Ly$\alpha$ wing we detected in these objects may be produced by a fraction of photons leaking through patchy dust distribution, and the luminosity of this component may be largest in Q01 and Q02, 
which have the highest bolometric luminosities among the sample.
Alternatively, the wing may be produced by radiative transfer of Ly$\alpha$ photons emitted from outside of the BLR; see \S \ref{sec:lya} for further discussion.
On the other hand, similarly luminous quasars with smaller extinctions ($A_{V, {\rm b}} < 2$) %
would be identified as unambiguous broad-line quasars with the rest-UV spectra, and would not be selected for the present study.
\citet{kato20} and \citet{iwamoto25} identified several candidate dust-reddened quasars with $A_V < 1$ in the SHELLQs sample, which may correspond to such objects that are absent from the present sample.

\begin{figure}
\epsscale{1.15}
\plotone{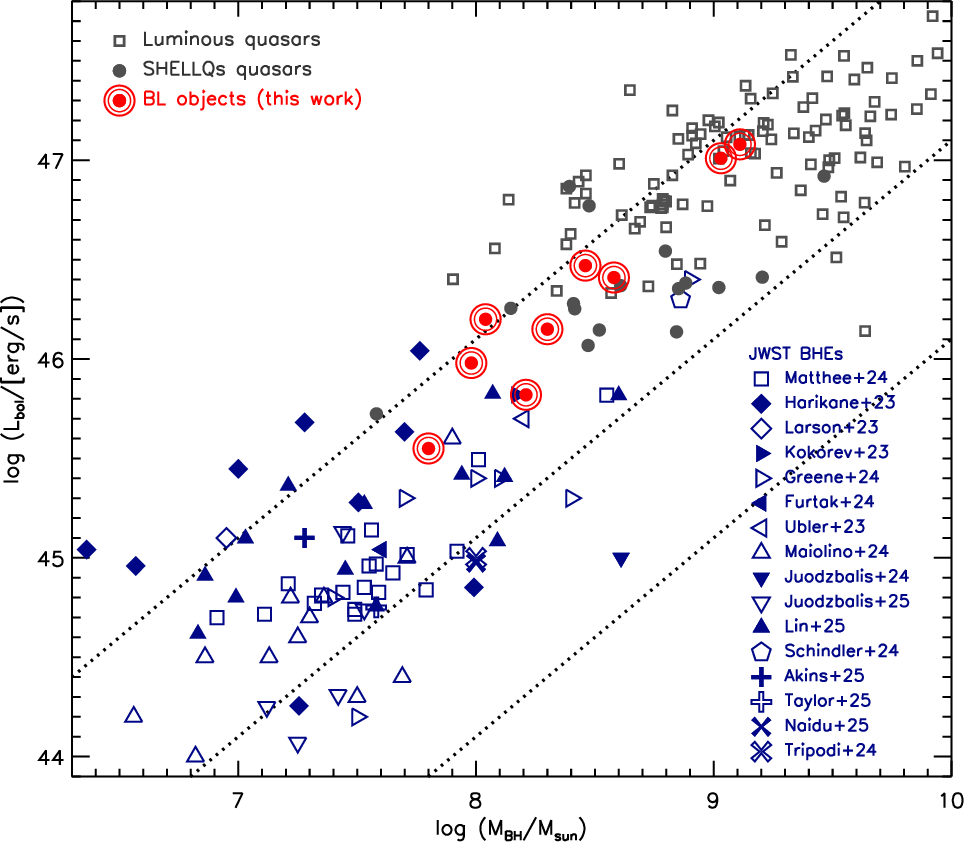}
\caption{
Black hole masses and bolometric luminosities of luminous quasars \citep[open squares;][]{willott10, shen19, yang21, yang23} and SHELLQs low-luminosity quasars 
\citep[grey dots;][M. Onoue et al., in prep.]{onoue19, p7, onoue24} at $z \sim 6 - 7$.
The dark blue symbols represent JWST BHEs at $z > 4$
\citep{harikane23,kokorev23,ubler23,furtak24,greene24,juodzbalis24,juodzbalis25,maiolino24,matthee24,schindler24,tripodi24,akins25,lin25,naidu25_lrd,taylor25}.
The dotted lines represent constant Eddington ratios of $\lambda_{\rm Edd}$ = 1, 0.1, and 0.01, from top to bottom.
\label{fig:agn_rel2}}
\end{figure}

Finally, we estimated the black hole mass ($M_{\rm BH}$) from the width and luminosity of the H$\alpha$ BLR component, following the recipe of \citet{greene05} and \citet{reines13}.
The resultant values ($M_{\rm BH} = 10^{7.8 - 9.1} M_\odot$) overlap with the distributions of low-$z$ SDSS quasars \citep[e.g.,][]{rakshit20, toba21} and UV-luminous quasars 
at $z \sim 6$ \citep[e.g.,][]{yang21}, but are at the lower end on average.
The Eddington ratios are in the range from $\lambda_{\rm Edd} = 0.3$ to 1.1, suggesting that sub-Eddington to Eddington accretion takes place in the BL objects. 
Figure \ref{fig:agn_rel2} compares the estimated values with those of classical quasars and other JWST BHEs,\footnote{
This figure is a collection of heterogeneous measurements, since different works used different assumptions and prescriptions to estimate 
$M_{\rm BH}$ and $L_{\rm bol}$. 
Dust extinction was corrected for in some papers, but not in all of them. 
For objects whose $L_{\rm bol}$ are not available, we estimated those values from broad H$\alpha$ luminosities (when they are available) with the bolometric correction of \citet{stern12}.
}
demonstrating that the present objects bridge the two AGN populations.
We also note the overall similarity of the present targets with the SDSS type-II quasar candidates at lower redshifts \citep{alexandroff13,greene14}, 
both of which are characterized by black hole masses of $M_{\rm BH} \sim 10^{8 - 9} M_\odot$ with sub-Eddington accretion, mild dust obscuration ($A_V \sim 0 - 3$), and 
high line luminosities ($L_{\rm [O III]} \sim 10^{43 - 45}$ erg s$^{-1}$).

However, we caution that our estimates reported above have large uncertainties.
Black hole masses derived from single-epoch spectra are associated with statistical uncertainty of 0.3 -- 0.5 dex \citep[e.g.,][]{vestergaard06}.
The bolometric correction to H$\alpha$ luminosity has been derived from low-$z$ AGNs, and we do not know how accurate it is when applied for the present sample of EoR objects.
We derived alternative estimates of bolometric luminosity from the continuum luminosity ($L_{5100}$) with the bolometric correction reported by \citet{shen11},
and from the [\ion{O}{3}] luminosity ($L_{\rm [O III]}$) with the bolometric correction of \citet{kauffmann09} and \citet{heckman14}.
Note that these corrections are not valid if the continuum or [\ion{O}{3}] emission is dominantly produced by starlight (which may well be the case; see the discussion in C. Phillips et al., in prep.), and also that the corrections may be inappropriate for the type of objects presented here. 
The results are listed in Table \ref{tab:alternative_Lbol} in Appendix \ref{sec:appendix}.
There is a broad agreement between the estimates from $L_{\rm H\alpha, b}$ and $L_{5100}$, while the $L_{\rm [O III]}$-based values are significantly higher.
When the $L_{\rm [O III]}$-based values are adopted, then all BL objects are accreting at super-Eddington rates.
This demonstrates 
the difficulty of applying correction factors derived from classical quasars and AGNs.

\subsection{Spatial extendedness} \label{sec:extendedness}

\begin{figure*}
\epsscale{1.15}
\plotone{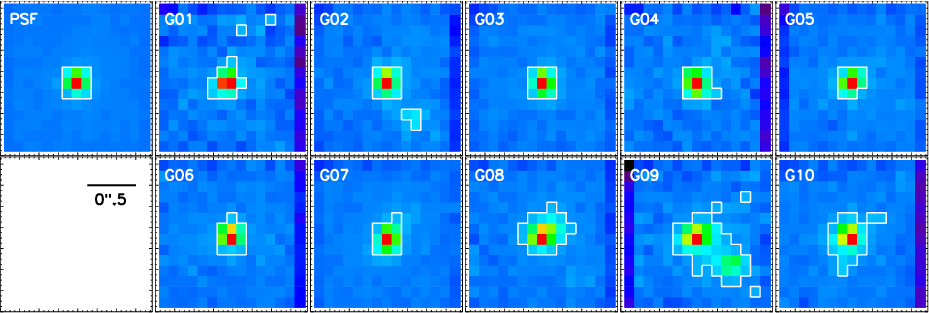}
\caption{
NIRSpec WATA images of G01 -- G10, compared with the PSF model. 
All the images have been flux-scaled consistently, in such a way that the central five pixels (the peak pixel and the adjacent four pixels) have the summed flux of unity.
Pixels with flux exceeding 5 \% of the peak value in each panel are outlined by the white lines. 
Each panel is 1".6 on a side. The bottom left panel provides a scale bar representing 0\arcsec.5, which corresponds to 2.9 kpc in proper distance at $z = 6.0$.
\label{fig:acq_images}}
\end{figure*}

The 13 objects are all spatially unresolved on the HSC-SSP images, with the median seeing of 0\arcsec.7 in the $z$ and $y$ bands \citep{aihara22}.
We observed three of them (G11, Q01, Q02) with JWST/NIRCam as part of GO 1967 in Cycle 1, using the F150W and F356W filters \citep{ding25}.
G11 is surrounded by prominent extended emission, which comes from the stellar populations within the host galaxy.
The decomposed host flux accounts for 60 -- 70 \% of the total flux of the system.
On the other hand, Q01 and Q02 have no or only marginal detection of extended emission, with the host to total flux ratios being 10 \% at most.

While the remaining objects have no NIRCam data, here we assess their spatial extendedness by exploiting the acquisition images of the NIRSpec observations.
The target acquisition was performed in the WATA mode, using the F140X filter.
This filter covers the wavelength from $\lambda_{\rm obs}$ = 0.8 to 2.0 $\mu$m, which is expected to contain the rest-UV continuum plus Ly$\alpha$ and other emission lines.
Each object was imaged in the 1\arcsec.6 $\times$ 1\arcsec.6 square S1600A1 aperture twice, with a total exposure time of 170 sec.
We applied spatial offset to the first image so that the peak pixel of the target matches that in the second image, and co-added them with a simple average.
For comparison, we created a PSF model\footnote{
NIRSpec undersamples the telescope resolution. The PSF model here includes the effect of the large angular pixel size (0.\arcsec1).}
 from the WATA F140X-band images of broad-line quasars, acquired as part of the GO1967 NIRSpec observations.
The images were processed in exactly the same way as for the present sample.

Figure \ref{fig:acq_images} (top left panel) shows the PSF model. 
The central pixel contains 40\% of the total flux, so the half-light diameter of the PSF is roughly 1 pixel (0.\arcsec1), which corresponds to 0.57 kpc at $z = 6.0$.
From the remaining panels, it appears that 
G01 -- G07 have a compact morphology comparable to the PSF, with most of the flux contained in the central 3 $\times$ 3 pixels.
On the other hand, G08 -- G10 are apparently extended.
If we assume that pixels with flux exceeding 5 \% of the peak value (below which contamination of background pixels become significant) belong to the source,
then the spatial extent of G08 -- G10 is roughly 0\arcsec.6 in diameter, while that of G01 -- G07 and PSF is 0\arcsec.3.
Combined with the Cycle 1 NIRCam measurements, it is immediately clear that the spatial extendedness is strongly tied to the presence/absence of broad lines.
All the nine BL objects (G01 -- G07, Q01, and Q02) have compact, almost unresolved morphology, while all the four NL objects (G08 -- G11) are accompanied by significant extended emission.
We will show below (\S \ref{sec:lrd}) that the contribution of emission lines is comparable to, or even larger than, the continuum in broadband photometry.
Indeed, the present objects have luminous Ly$\alpha$ emission in the F140X band, as confirmed by the SHELLQs ground-based spectroscopy.
Thus the unresolved emission of the BL objects may be due to strong lines from nuclear regions, while the extended emission of the NL objects come from much larger spatial scales of the galaxies.
Further analyses of extended emission around the present targets, exploiting the 2d spectra, will be presented in C. Phillips et al. (in prep.).

Interestingly, G02 and G09 are accompanied by a nearby source at distance of $\sim$0\arcsec.5.
No counterpart sources are detected on the HSC optical images, which could be due to the limited resolution and sensitivity. 
Similar discoveries of companion galaxies have been reported around EoR quasars, some of which have been spectroscopically confirmed 
\citep[e.g.,][]{decarli17, decarli19a, decarli19b, p20}.
If the nearby sources in Figure \ref{fig:acq_images} are physically associated companions, then it may suggest that the objects presented here live in similarly rich galaxy environments to quasars.
Future deep observations with sub-arcsec resolution, such as those with JWST and ALMA, are needed to further investigate this issue.

\begin{figure*}
\epsscale{1.0}
\plotone{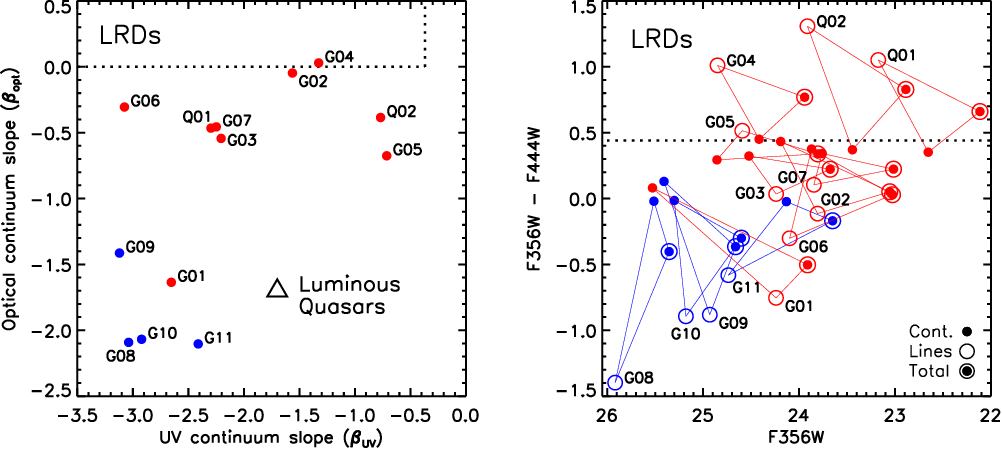}
\caption{
The continuum slopes (left panel) and synthesized broadband magnitudes and colors (right panel) of the BL (red symbols) and NL (blue symbols) objects.
The dotted lines represent the selection criteria of LRDs used by \citet{kocevski24}.
The triangle shows continuum slopes of classical luminous quasars \citep{selsing16}.
The dots and open circles in the right panel represent magnitudes for a power-law fit to continuum and for all the combined emission lines, respectively, while
the dots with open circles represent their sum. 
Measurements of the same object are connected by the solid lines.
\label{fig:bb_mags}}
\end{figure*}

\subsection{Are they LRDs?} \label{sec:lrd}

As we described in \S \ref{sec:intro}, a significant fraction of JWST BHEs belong to a galaxy population called LRDs, which are 
distinguished by compact shape and V-shaped spectra from the rest-frame UV to optical wavelength.
The discovery of this abundant population is one of the biggest surprises that JWST has made in its first few years of operation \citep[e.g.,][]{onoue23,kocevski24,matthee24}.
Here we attempt to address how the present sample of 13 objects is related to this class of galaxies, though the available data are rather limited.

The definition of LRDs differs slightly from paper to paper.
We adopt the one used by \citet{kocevski24}, whose criteria are given with the spectral slopes $\beta_{\rm UV} < -0.37$ and $\beta_{\rm opt} > 0$.
Here the slopes are defined such that $f_\lambda \propto \lambda^{\beta_{\rm UV}}$ and $\propto \lambda^{\beta_{\rm opt}}$ 
at the blue and red side of the Balmer break \citep[$\lambda_{\rm rest}$ = 3645 \AA;][]{setton24}, respectively.
We estimated the UV slope from the continuum fluxes measured at redward of Ly$\alpha$ ($\lambda_{\rm rest}$ $\sim$ 1300 \AA) and at $\lambda_{\rm rest}$ = 3645 \AA.
The former was available from the SHELLQs discovery spectra, while the latter was calculated with the best-fit continuum model described in \S \ref{sec:specmodel}.
Note that both the discovery spectra and the present NIRSpec spectra have been corrected for signal loss due to finite slit widths.
The slope $\beta_{\rm opt}$ was taken directly from the best-fit power-law continuum model.

As is clear from Figure \ref{fig:bb_mags} (left panel), all the objects have sufficiently blue rest-UV continuum, meeting one of the LRD conditions: $\beta_{\rm UV} < -0.37$.
Among them, one object (G04) meets the other LRD condition: $\beta_{\rm opt} > 0$.
Several others have fairly flat rest-optical continuum with $\beta_{\rm opt} > -0.5$, and are relatively close to the borderline between LRDs and non-LRDs.
Interestingly, the BL objects have apparently larger $\beta_{\rm opt}$ than do the NL objects.
Such strong connection between the continuum slope and the presence/absence of broad lines may indicate that the continuum emission is contributed significantly by AGN radiation, at least in the BL objects.
It may then partially explain why the BL objects are exclusively unresolved in the WATA images (see Figure \ref{fig:acq_images}).

\begin{figure*}
\epsscale{1.0}
\plotone{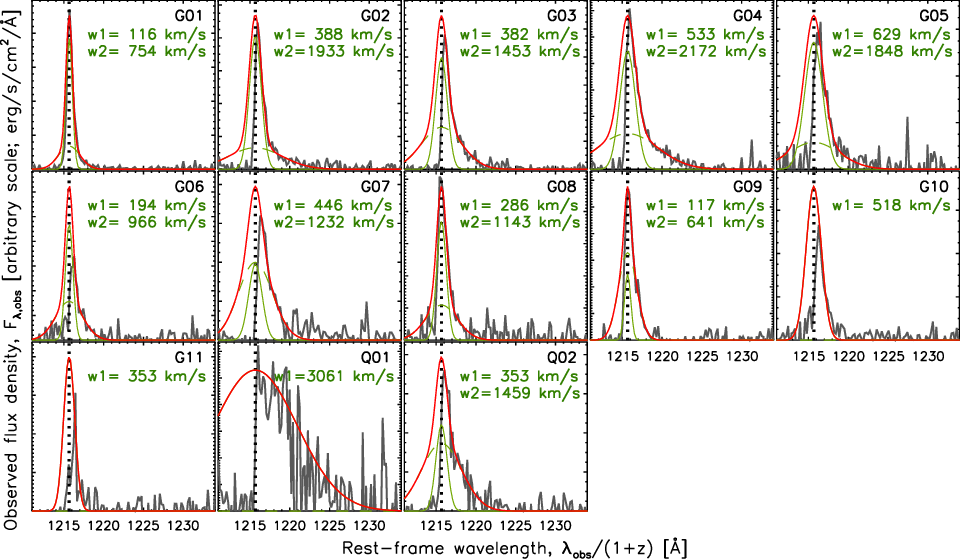}
\caption{
The rest-UV spectra around Ly$\alpha$, taken from the SHELLQs discovery papers (see Table \ref{tab:targets} for the reference). 
The red lines represent the fitted models, composed of at most two Gaussian functions (green solid and dashed lines).
The fitting was performed at wavelengths redward of the observed line peak (see text).
The FWHMs of the Gaussians are reported at the top right corner of each panel; note that these values refer to the intrinsic widths free from the IGM absorption,
and are thus larger than those listed in Table \ref{tab:targets}.
The dotted lines mark the expected central wavelengths of Ly$\alpha$, given the systemic redshifts.
\label{fig:spec_lya}}
\end{figure*}

On the other hand, many previous studies used broadband photometry to select LRDs \citep[e.g.,][]{barro24, greene24, labbe25}.
Since our NIRSpec spectra show that the emission lines have enormous EWs, we could ask whether the present targets would be identified as LRDs with such selection.
In Figure \ref{fig:bb_mags} (right panel), we present the synthesized photometry with the NIRCam F356W and F444W bands, which are covered by the NIRSpec spectra.
In addition to the total magnitudes, we calculated magnitudes separately for power-law continuum and for all the combined emission lines
from the best-fit spectral models.
These calculations demonstrate that the contribution of the emission lines is comparable to, or even larger than, the continuum in the broadbands.\footnote{
We expect a similar situation in the NIRSpec F140X band (used in \S \ref{sec:extendedness}), which contains strong Ly$\alpha$ and \ion{C}{4} $\lambda$1549 lines.}
The rest-optical LRD criterion ($\beta_{\rm opt} > 0$) corresponds to the magnitude difference of $F356W - F444W > 0.44$, which is satisfied by G04, Q01, and Q02.
That is, these three objects would be selected as LRD candidates from broadband photometry alone.
The latter two objects may be among the first previously-known quasars that are plausible LRD candidates \citep[e.g.,][]{stepney24}.

It is worth noting that G01 is an outlier among the BL objects in several aspects.
Figure \ref{fig:bb_mags} shows that its optical continuum slope is closer to the NL objects than to other BL objects.
It also has the smallest broad-to-narrow component ratio of the H$\alpha$ flux among the BL objects
(see Table \ref{tab:ew_av}).
Its narrow [\ion{O}{3}] and H$\alpha$ lines have very high EWs, comparable to extreme emission-line galaxies \citep[e.g.,][]{boyett24, llerena24}.
Finally, it has a hint of weak extended emission in the NIRSpec WATA image (Figure \ref{fig:acq_images}).
These facts suggest that G01 has a large flux contribution from star formation in the host galaxy, compared to other BL objects which may be dominated by AGN emission.

\subsection{Revisiting Ly$\alpha$ profiles} \label{sec:lya}

Given the detection of broad permitted lines in many of the targets, here we revisit their Ly$\alpha$ profiles in the rest-UV spectra, taken from the discovery papers \citep[e.g.,][]{p2}.
Our previous analysis was quite limited in the SHELLQs sample, due to the lack of accurate systemic redshifts for most of the objects\footnote{
It turned out that the Ly$\alpha$ redshifts are close to the systemic redshifts in the present sample, with prominent Ly$\alpha$ lines.
The difference of the two redshifts ($z_{\rm Ly\alpha} - z_{\rm sys}$) ranges from $-0.003$ to $+0.005$.
}
and relatively low S/N of the spectra.
As presented in Figure \ref{fig:spec_lya}, we found that many of the targets show a weak extended wing in Ly$\alpha$, at the base of a narrow core component. 
We performed a crude line fitting with up to two Gaussian components, as shown in the figure.
The central wavelengths of all Gaussians were fixed to the values expected from the systemic redshifts.
The fitting was performed at wavelengths redward of the observed line peak (i.e., not the central wavelength), since the blue side is heavily affected by \ion{H}{1} absorption from the IGM.

We found that only one of the four NL objects (G08 -- G11) has a wing with FWHM exceeding 1000 km s$^{-1}$.
It is consistent with the lack of broad component in the rest-optical permitted lines. 
On the other hand, Q01 is dominated by broad Ly$\alpha$, and Q02 has a substantial broad component.
These two objects were classified into quasars due to these features in the discovery papers.
Most of the remaining BL objects (G01 -- G07) have Ly$\alpha$ wings, but their FWHMs 
are generally smaller than those of the BLR H$\alpha$ emission (see Table \ref{tab:ew_av}).

The rough agreement on the presence/absence of broad component with other permitted lines may indicate that some of the broad Ly$\alpha$ emission originates from the BLR. 
However, such an interpretation may be premature, since the radiative transfer of Ly$\alpha$ photons is very complicated.
Due to efficient resonant scattering within the emitting gas clouds, which may be non-static and also cause dust extinction, the 
Ly$\alpha$ line profile can be substantially altered before escaping the galaxy.
In extreme cases, the emergent line profile shows multiple peaks, and the profile can extend to $>$1000 km s$^{-1}$ from the original central wavelength
 \citep[e.g.,][]{songaila18, verhamme06, verhamme08, kulas12, matthee18}.
The observed Ly$\alpha$ profile of our objects can be significantly affected by such complicated processes.

\subsection{AGN fraction and number density}

The targets of this work are a rare class of galaxies with extremely luminous and narrow Ly$\alpha$ emission. 
In particular, the JWST GO 3417 program was designed to observe all the SHELLQs galaxies with $L_{\rm Ly\alpha} \ge 10^{43.8}$ erg s$^{-1}$ (G01 -- G10) known at the time we wrote the proposal.
We detected broad permitted lines in seven of them, indicating at face value an AGN fraction of $f_{\rm AGN}^\prime = 0.7$. 
Here we estimate the true AGN fraction among such luminous LAEs (LLAEs, which refers to objects with $L_{\rm Ly\alpha} \ge 10^{43.8}$ erg s$^{-1}$ hereafter), 
by considering possible selection effects.\footnote{
We clarify that the three other objects discussed in this paper are excluded from the argument of this section, since their selection for the JWST GO 1967 observations was not primarily based on Ly$\alpha$ luminosity (M. Onoue et al., in prep.).
G11 is a less luminous LAE with $L_{\rm Ly\alpha} = 10^{43.3}$ erg s$^{-1}$, and Q01 and Q02 are quasars. 
}

\begin{figure}
\epsscale{1.15}
\plotone{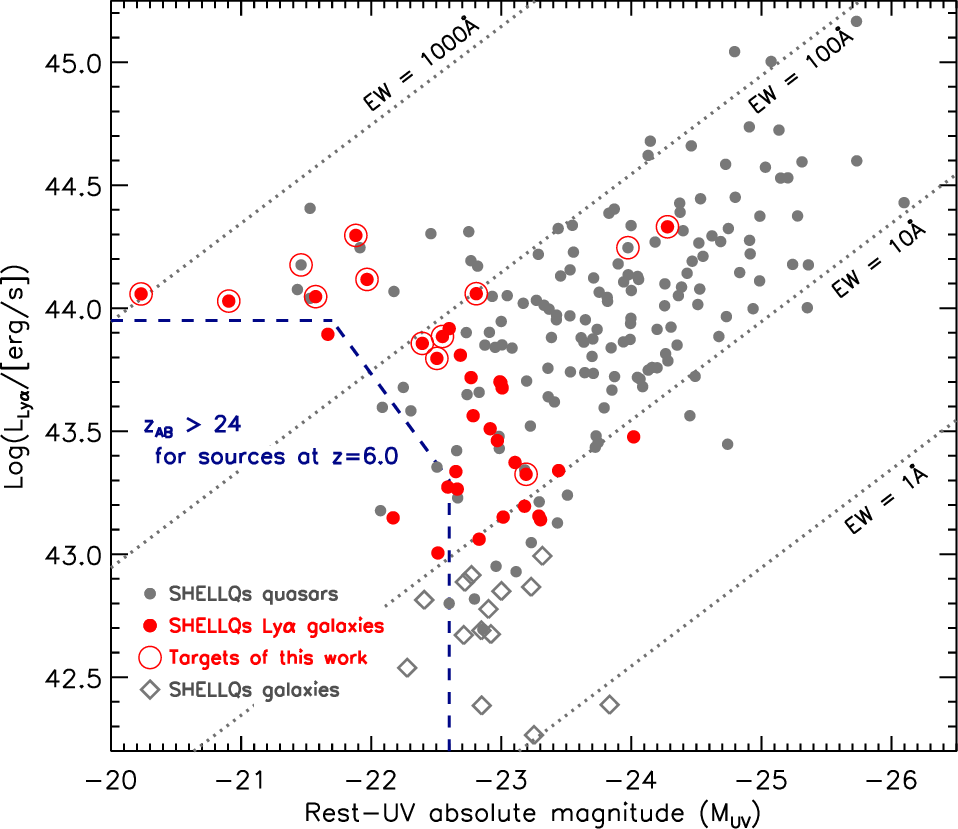}
\caption{
Rest-UV absolute magnitudes and Ly$\alpha$ luminosities of the SHELLQs quasars (gray dots), galaxies with (red dots) and without (diamonds) luminous Ly$\alpha$ lines.
The dotted lines represent constant rest-frame EWs of 1, 10, 100, and 1000 \AA.
Sources to the left and bottom of the dashed lines are fainter than our primary limiting magnitude ($z_{\rm AB} = 24.0$) at $z = 6.0$.
\label{fig:select_func}}
\end{figure}

The SHELLQs target selection for follow-up spectroscopy is mainly based on source magnitudes, colors, and spatial extendedness.
Figure \ref{fig:select_func} presents the rest-UV absolute magnitudes and Ly$\alpha$ luminosities of spectroscopically-identified SHELLQs objects at $z \ga 6$.
The deficit of objects at the bottom left corner is due to our primary magnitude limit, $z_{\rm AB} < 24$.
While $z = 6$ galaxies with $L_{\rm Ly\alpha} > 10^{44}$ erg s$^{-1}$ can meet this limit with Ly$\alpha$ flux alone,
those with smaller $L_{\rm Ly\alpha}$ need additional continuum flux in order to be selected.
Given that the BL objects have higher line EWs (i.e., smaller continuum contribution) than the NL objects (see Table \ref{tab:ew_av}), selection with broadband magnitudes biases against AGNs for a given Ly$\alpha$ luminosity.
That is, the AGN fraction among LLAEs would be higher than $f_{\rm AGN}^\prime = 0.7$ if the magnitude limit were relaxed.
Figure \ref{fig:select_func} also demonstrates that the Ly$\alpha$ EWs of LLAEs are much higher than quasars on average.
It leads to very blue $z - y$ colors for LLAEs at $z \sim 6$, placing them far away from contaminating brown dwarfs in the color space and thus making them easy to select as quasar candidates
\citep[see, e.g., Fig. 1 of][]{p2}.
Hence the selection probability is essentially 100 \% for LLAEs, based on their colors alone. 

On the other hand, the selection of LLAEs is largely affected by the condition on spatial extendedness.
While SHELLQs selected both unresolved and resolved sources from the HSC images, higher priority was given to the former in follow-up spectroscopy.
All the 13 objects in this paper are unresolved at HSC resolution (see \S \ref{sec:extendedness}).
In order to assess the associated selection effects,
we make use of a sample from the HSC-based SILVERRUSH project \citep{ouchi18, shibuya18}.
This project identified a large number of LAEs at multiple redshifts, by exploiting a rich set of narrow band filters installed on HSC \citep{kawanomoto18, inoue20}.
We use the sample at $z = 6.6$, whose Ly$\alpha$ emission is covered by the same HSC $z$ band as for our targets (except for the one at $z = 6.88$, whose Ly$\alpha$ is covered by the $y$ band).
We found that the fraction of LAEs meeting our unresolved criterion 
\citep[0.7 $<$ $\mu$/$\mu_{\rm PSF}$ $<$ 1.2, where $\mu$ and $\mu_{\rm PSF}$ represent 
the image adaptive moment of a given source and PSF model, respectively;][]{p10} is 30 \%.\footnote{
The SILVERRUSH LAEs used here have lower Ly$\alpha$ luminosities ($L_{\rm Ly\alpha} = 10^{43.0 - 43.7}$ erg s$^{-1}$) than the present 10 objects.
Given that the AGN fraction (and hence unresolved fraction) is larger at higher $L_{\rm Ly\alpha}$ \citep{konno16, sobral18, spinoso20}, we expect that the fraction of unresolved LAEs at $L_{\rm Ly\alpha} > 10^{43.8}$ erg s$^{-1}$ is actually higher than 30 \%.
}
Given the AGN fraction of $f^\prime_{\rm AGN} = 0.7$ in the unresolved LLAEs, the lower limit of the true AGN fraction is $f_{\rm AGN}^{\rm min} = 0.2$; it is obtained by assuming 
the extreme case in which none of the resolved LLAEs hosts an AGN.

\begin{figure}
\epsscale{1.15}
\plotone{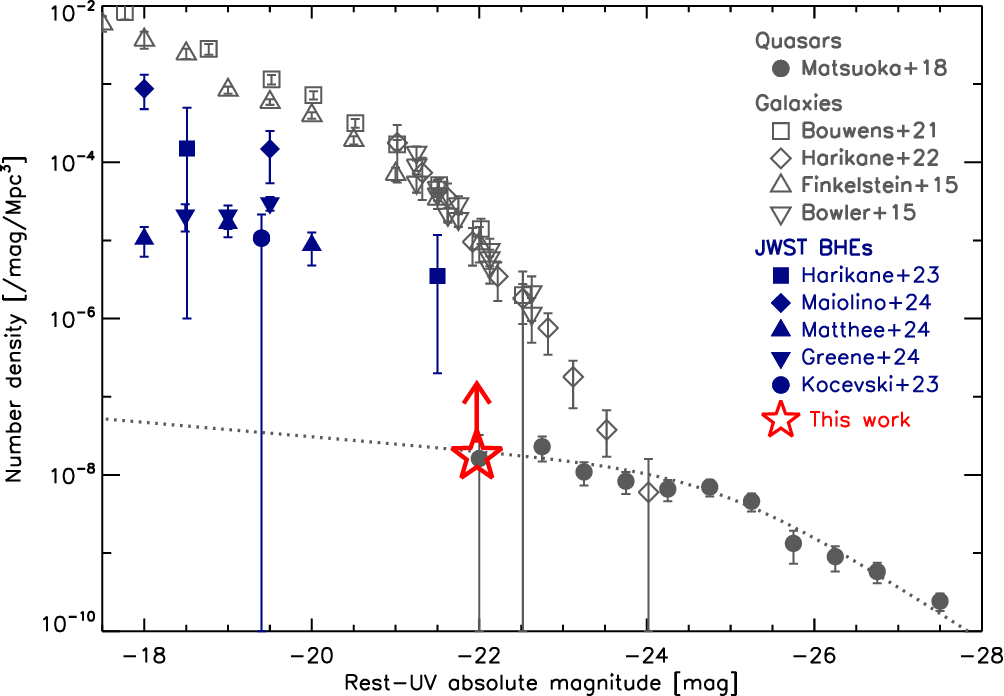}
\caption{
The luminosity functions of classical quasars at $z = 6$ \citep[gray dots and dotted line;][]{p5} and of galaxies at $z \sim 6$ \citep{bowler15, finkelstein15, bouwens21, harikane22}, 
compared with the number densities of JWST BHEs at $z \sim 4 - 6$ \citep{harikane23,kocevski23,greene24,maiolino24,matthee24}.
The star represents the lower limit of AGN number density derived in this work. 
\label{fig:density}}
\end{figure}

Based on the latest luminosity function from SILVERRUSH \citep{umeda24}, the number densities of LAEs with $L_{\rm Ly\alpha} > 10^{43.8}$ erg s$^{-1}$ 
are $2 \times 10^{-7}$ and $8 \times 10^{-8}$ Mpc$^{-3}$ at $z = 5.7$ and $z = 6.6$, respectively.
At the median redshift of our objects ($z_{\rm med} = 6.1$), which is close to the midpoint of the above two redshifts, we estimate the LLAE number density to be roughly
$1 \times 10^{-7}$ Mpc$^{-3}$. 
With the minimum AGN fraction $f_{\rm AGN}^{\rm min} = 0.2$ and the $M_{\rm UV}$ range of the sample quantified by twice the median absolute deviation,
we get $2 \times 10^{-8}$ Mpc$^{-3}$ mag$^{-1}$ as the lower limit of AGN number density. 
It is plotted in Figure \ref{fig:density} at the median magnitude of our sample, $M_{\rm UV} = -22.0$.
The AGN number density at this magnitude could be significantly higher, since (i) selection with the broadband magnitude limit biases against AGNs in LLAEs, (ii) there may be AGNs among resolved LLAEs, and (iii) LAEs with $L_{\rm Ly\alpha} < 10^{43.8}$ erg s$^{-1}$ at similar $M_{\rm UV}$ are not counted.

Figure \ref{fig:density} also presents number densities of classical quasars and galaxies at $z \sim 6$, and other BHEs at $z \sim 4 - 6$ taken from the literature.
Our lower limit is a few orders of magnitude lower than the number densities of other BHEs, which implies that the present sample represents a rare class of BHEs with the highest continuum and line luminosities.
On the other hand, the population represented by our sample is at least as numerous as unobscured quasars with the same continuum luminosity.
It suggests that a substantial fraction of active SMBHs have been overlooked in the EoR, even in the luminosity range accessible with previous high-$z$ quasar surveys.
Since our objects are moderately dust obscured and appear as normal galaxies in the rest-UV wavelengths (except for their extremely high Ly$\alpha$ luminosity, which is not usually considered as conclusive evidence for AGN), ground-based observations alone have been inadequate to reveal their AGN.
Future JWST observations of a wider population of EoR galaxies, e.g., resolved LLAEs or LAEs with lower luminosity, 
will provide further clues to understanding the whole picture of SMBH activity during cosmic dawn.

\section{Summary} \label{sec:summary}

We present the results from JWST/NIRSpec observations of 13 objects in the EoR.
The sample was originally selected by the HSC-SSP imaging survey over $>$1000 deg$^2$, and all the objects have luminous Ly$\alpha$ emission ($L_{\rm Ly\alpha} > 10^{43}$ erg s$^{-1}$) in the rest-UV spectra.
Among them, 10 galaxies (G01 -- G10) were selected for their high Ly$\alpha$ luminosities for Cycle-2 observations in GO 3417.
The remaining three objects include a similar galaxy (G11) and two quasars with relatively narrow Ly$\alpha$ lines (Q01 and Q02), which were observed as part of GO 1967 in Cycle 1.

We found that the majority of the sample (G01 -- G07, Q01, and Q02; BL objects) exhibit a broad component in the rest-optical \ion{H}{1} Balmer lines and \ion{He}{1} lines.
The FWHMs of the broadest component range from 2000 to 5000 km s$^{-1}$, which we interpret as a signature of BLR emission originating from AGNs.
We found a faint extended wing component in the Ly$\alpha$ line of these objects from our ground-based spectra, but its origin is unclear, due to the complicated radiative transfer of resonant Ly$\alpha$ photons.
The remaining four objects (G08 -- G11; NL objects) show strong [\ion{O}{3}] and other high-ionization lines.
However, they are typical of EoR galaxies discovered by JWST, and the nature of these objects remains unclear.
Most of the 13 objects are characterized by mild dust obscuration with $0 < A_V < 3$.
The intrinsic bolometric luminosities estimated for the BL objects amount to $L_{\rm bol} = 10^{45.6 - 47.1}$ erg s$^{-1}$, suggesting that they represent the long-sought UV-obscured counterparts of luminous quasars in the EoR.
They host SMBHs with $M_{\rm BH} = 10^{7.8 - 9.1} M_\odot$, undergoing sub-Eddington to Eddington accretion. 

From analysis of the NIRSpec target acquisition images for the Cycle 2 targets,  
we found that all the NL objects have extended emission, with a diameter of roughly 0\arcsec.6.
On the other hand, all the BL objects are consistent with being spatially unresolved, with the half-light diameter less than around 0\arcsec.1 (corresponding to 0.57 kpc at $z = 6.0$).
Combined with the NIRCam analysis of the Cycle 1 targets \citep{ding25}, 
it provides an additional support for the presence of an AGN in the BL objects, whose strong line (and possibly continuum) emission comes from the unresolved nucleus.
We found that a few (G04, and possibly Q01, Q02) of the BL objects have rest-UV and rest-optical colors meeting the LRD definition, and that several others have colors close to the boundary between LRDs and non-LRDs.

GO 3417 was designed to observe all the SHELLQs galaxies with $L_{\rm Ly\alpha} > 10^{43.8}$ erg s$^{-1}$ (LLAEs) known at the time we wrote the proposal.
By considering possible selection effects, we estimate that the AGN fraction among similarly luminous LAEs in the EoR is no less than $f_{\rm AGN}^{\rm min} = 0.2$. 
With this fraction and the LAE luminosity function taken from the HSC-based SILVERRUSH project, we get 
$2 \times 10^{-8}$ Mpc$^{-3}$ mag$^{-1}$ as the lower limit of AGN number density, at the continuum magnitude of $M_{\rm UV} = -22.0$.
This density is comparable to the number density of classical unobscured quasars at similar redshifts and magnitudes, 
suggesting that a substantial fraction of active SMBHs have been overlooked in the past rest-UV surveys for EoR quasars.

Our JWST observations have revealed
that the most luminous galaxies with strong Ly$\alpha$ emission are hosts of obscured quasars in the EoR.
The observed properties indicate that they represent {\it the highest-luminosity family of BHEs}, with colors close to those of LRDs. 
On the other hand, the SHELLQs project has discovered a number of unobscured low-luminosity quasars (see Figure \ref{fig:agn_rel}), which represent {\it the lowest-luminosity family of UV-luminous quasars} discovered over the past few decades.
Taken together, the SHELLQs sample may represent a key population that bridges the gap between classical quasars and newly emerged BHEs/LRDs.
We have an approved JWST Cycle 4 program (GO 7491, PI: Y. Matsuoka) to observe an additional 30 SHELLQs galaxies in the EoR, including those with weak Ly$\alpha$ emission, which may provide further evidence to link the two distinct AGN populations.

\acknowledgments

We thank the anonymous referee for their constructive feedback, which helped improve the quality of this paper.

This work is based on observations made with the NASA/ESA/CSA James Webb Space Telescope. 
The data were obtained from the Mikulski Archive for Space Telescopes (MAST) at the Space Telescope Science Institute (STScI), which is operated by the Association of Universities for Research in Astronomy, Inc., under NASA contract NAS 5-03127 for JWST. 
These observations are associated with programs GO 1967 and 3417.

The data described here may be obtained from the MAST archive at
\dataset[doi:10.17909/6vpm-5k66]{https://dx.doi.org/10.17909/6vpm-5k66}.
The data available there, however, were pipeline-processed at the STScI, and are therefore not identical to those presented in this paper.

This research is based in part on data collected at Subaru Telescope, which is operated by the National Astronomical Observatory of Japan. 
We are honored and grateful for the opportunity of observing the Universe from Maunakea, which has the cultural, historical and natural significance in Hawaii.
We appreciate the staff members of the telescope for their support during our FOCAS observations.

This work is based in part on observations made with the GTC, installed at the Spanish Observatorio del Roque de los Muchachos of the Instituto de Astrof\'{i}sica de Canarias, on the island of La Palma.

Y. M. was supported by the Japan Society for the Promotion of Science (JSPS) KAKENHI Grant No. 21H04494. 
K. I. acknowledges the support under the grant PID2022-136827NB-C44 provided by MCIN/AEI /10.13039/501100011033 / FEDER, EU.
J. D. S. is supported by the JSPS KAKENHI Grant No. JP22H01262 and the World Premier International Research Center Initiative (WPI), MEXT, Japan.
K. K. acknowledges the support by JSPS KAKENHI Grant Numbers 22H04939, 23K20035, and 24H00004.

\clearpage

\appendix

\section{Spectral measurements} \label{sec:appendix}

The emission line fluxes and continuum properties of the 13 objects were measured from the best-fit spectral models, and are reported in the observer's frame in Tables \ref{tab:line_fluxes1} and \ref{tab:line_fluxes2}.
Table \ref{tab:alternative_Lbol} lists the bolometric luminosities and Eddington ratios derived from the continuum luminosities at 5100 \AA\ ($L_{5100}$) and from [\ion{O}{3}] $\lambda$5007 luminosities.

\begin{longrotatetable}
\begin{deluxetable*}{ccccccccccccccccccc}
\tablecaption{Emission line fluxes and continuum properties of G01 -- G10 \label{tab:line_fluxes1}}
\tablewidth{700pt}
\tabletypesize{\scriptsize}
\tablehead{
\colhead{} &                                              \colhead{G01} &      \colhead{G02} &        \colhead{G03} &      \colhead{G04} &       \colhead{G05} &        \colhead{G06} &       \colhead{G07} &        \colhead{G08} &       \colhead{G09} &        \colhead{G10}   
} 
\startdata
\hline\multicolumn{11}{c}{SF/NLR lines}\\\hline
                                                   H9 &   \nodata               &    \nodata               &    \nodata               &    \nodata               &    \nodata               &   1.53 $\pm$  0.10 &    \nodata               &    \nodata               &    \nodata               &    \nodata               \\
                                                   H8 &   \nodata               &    \nodata               &    \nodata               &    \nodata               &    \nodata               &   1.98 $\pm$  0.12 &    \nodata               &    \nodata               &    \nodata               &    \nodata               \\
                                     H$\epsilon$ &   \nodata               &   2.66 $\pm$  0.12 &    \nodata               &    \nodata               &    \nodata               &   2.84 $\pm$  0.11 &    \nodata               &   0.63 $\pm$  0.18 &    \nodata               &    \nodata               \\
                                        H$\delta$ &  3.61 $\pm$  0.17 &   2.99 $\pm$  0.12 &   3.14 $\pm$  0.22 &   1.09 $\pm$  0.12 &   2.51 $\pm$  0.14 &   3.87 $\pm$  0.11 &   3.22 $\pm$  0.15 &   0.78 $\pm$  0.15 &   1.97 $\pm$  0.25 &    \nodata               \\
                                   H$\gamma$  &  6.15 $\pm$  0.15 &   5.25 $\pm$  0.13 &   7.24 $\pm$  0.20 &   2.19 $\pm$  0.10 &   4.35 $\pm$  0.15 &   6.82 $\pm$  0.13 &   5.33 $\pm$  0.16 &   1.09 $\pm$  0.13 &   3.21 $\pm$  0.27 &   3.07 $\pm$  0.16 \\
                                         H$\beta$ & 14.62 $\pm$  0.19 & 12.01 $\pm$  0.23 & 16.25 $\pm$  0.26 &   5.33 $\pm$  0.14 &   8.51 $\pm$  0.24 &    \nodata              & 12.70 $\pm$  0.31 &   2.44 $\pm$  0.20 &   7.25 $\pm$  0.56 &   6.70 $\pm$  0.25 \\
                                       H$\alpha$ & 43.32 $\pm$  0.41 & 40.61 $\pm$  0.67 & 65.66 $\pm$  0.76 &  28.89 $\pm$  0.55 &  26.95 $\pm$  0.72 & 50.71 $\pm$  0.50 & 43.33 $\pm$  0.98 & 6.74 $\pm$  0.50 &  23.75 $\pm$  1.80 &  20.08 $\pm$  0.71 \\
             \ion{He}{1} $\lambda$4473 &  0.78 $\pm$  0.10 &   0.81 $\pm$  0.08 &   0.91 $\pm$  0.14 &   0.36 $\pm$  0.07 &   0.82 $\pm$  0.09 &   1.04 $\pm$  0.09 &   0.95 $\pm$  0.09 &   $<$0.14               &   $<$0.51                &   $<$0.52               \\
             \ion{He}{2} $\lambda$4687 &  0.36 $\pm$  0.09 &   0.89 $\pm$  0.08 &   1.07 $\pm$  0.13 &   0.39 $\pm$  0.07 &   2.10 $\pm$  0.10 &   0.65 $\pm$  0.09 &   $<$0.07               &   $<$0.14               &   $<$0.52                &   0.31 $\pm$  0.10 \\
             \ion{He}{1} $\lambda$5877 &  2.07 $\pm$  0.14 &   2.62 $\pm$  0.12 &   3.85 $\pm$  0.18 &   1.51 $\pm$  0.11 &   3.69 $\pm$  0.15 &   3.90 $\pm$  0.16 &   4.48 $\pm$  0.16 &   0.53 $\pm$  0.13 &   1.18 $\pm$  0.16 &   0.49 $\pm$  0.13 \\
             \ion{He}{1} $\lambda$6679 &  $<$0.86               &   $<$0.93              &   1.21 $\pm$  0.25 &   $<$0.75               &   1.68 $\pm$  0.17 &    \nodata               &   1.52 $\pm$  0.17 &   $<$0.28                &   $<$0.81                &   $<$0.95               \\
             \ion{He}{1} $\lambda$7067 &  2.42 $\pm$  0.23 &    \nodata               &   4.61 $\pm$  0.29 &   1.65 $\pm$  0.18 &   4.73 $\pm$  0.25 &    \nodata               &   5.44 $\pm$  0.24 &   $<$0.89               &   $<$0.95               &   0.71 $\pm$  0.23 \\
   $[$\ion{Ne}{3}$]$ $\lambda$3869 &  \nodata                &    \nodata               &    \nodata               &    \nodata               &    \nodata               &   8.06 $\pm$  0.15 &    \nodata               &    \nodata               &    \nodata               &    \nodata               \\
    $[$\ion{O}{3}$]$ $\lambda$4363 &  3.10 $\pm$  0.12 &   3.72 $\pm$  0.11 &   5.41 $\pm$  0.17 &   2.54 $\pm$  0.12 &   1.85 $\pm$  0.13 &   4.36 $\pm$  0.10 &   0.69 $\pm$  0.13 &   0.41 $\pm$  0.10  &   0.78 $\pm$  0.14 &   1.09 $\pm$  0.12 \\
    $[$\ion{O}{3}$]$ $\lambda$5007 & 94.48 $\pm$  0.51 & 97.98 $\pm$  0.92 & 89.81 $\pm$  0.81 & 38.57 $\pm$  0.41 & 45.33 $\pm$  0.70 & 94.93 $\pm$  0.38 & 108.60 $\pm$  1.43 & 19.25 $\pm$  0.73 & 54.56 $\pm$  3.36 & 43.08 $\pm$  0.90 \\
    $[$\ion{O}{1}$]$ $\lambda$6300 &  $<$0.55                &   0.76 $\pm$  0.15 &   $<$1.00              &   0.63 $\pm$  0.15 &   $<$0.62               &   $<$0.28              &   0.65 $\pm$  0.15 &   $<$0.74                  &   $<$0.86               &   $<$0.42                \\
    $[$\ion{N}{2}$]$ $\lambda$6585 &  $<$0.11                &   $<$0.77               &   $<$1.19              &   0.80 $\pm$  0.18 &   0.82 $\pm$  0.27 &   $<$1.27               &   0.91 $\pm$  0.25 &   $<$0.18                 &   1.38 $\pm$  0.19 &   0.59 $\pm$  0.17 \\
    $[$\ion{S}{2}$]$ $\lambda$6716 &  $<$0.10                &   $<$0.11               &   $<$0.16              &   $<$0.27               &   $<$1.05               &    \nodata               &   0.71 $\pm$  0.19 &   $<$0.38                  &   1.00 $\pm$  0.19 &   $<$0.74                 \\
    $[$\ion{S}{2}$]$ $\lambda$6731 &  $<$0.32                &   $<$0.22               &   $<$0.76              &   $<$0.47               &   $<$0.98               &    \nodata               &   $<$0.88               &   $<$0.61                 &   0.76 $\pm$  0.20 &   $<$1.02                 \\
\hline\multicolumn{11}{c}{BLR lines}\\\hline
                                                   H9 &   \nodata               &    \nodata               &    \nodata               &    \nodata               &    \nodata               &   0.77 $\pm$  0.06 &    \nodata               &    \nodata                 &   \nodata                 &   \nodata                 \\
                                                   H8 &   \nodata               &    \nodata               &    \nodata               &    \nodata               &    \nodata               &   3.42 $\pm$  0.56 &    \nodata               &    \nodata                 &   \nodata                 &   \nodata                 \\
                                    H$\epsilon$ &   \nodata               &   3.36 $\pm$  0.57 &    \nodata               &    \nodata               &    \nodata               &   2.19 $\pm$  0.53 &    \nodata               &    \nodata                 &   \nodata                 &   \nodata                 \\
                                       H$\delta$ &  $<$3.07                &   4.70 $\pm$  0.47 &   $<$3.90              &   3.56 $\pm$  0.69 &   3.75 $\pm$  0.67 &   3.91 $\pm$  0.50 &   6.05 $\pm$  0.51 &    \nodata                 &   \nodata                 &   \nodata                 \\
                                   H$\gamma$ &  $<$1.70               &   7.92 $\pm$  0.45 &   3.61 $\pm$  0.56 &   6.78 $\pm$  0.55 &   5.29 $\pm$  0.59 &   7.94 $\pm$  0.52 &   9.84 $\pm$  0.45 &    \nodata                 &   \nodata                 &   \nodata                 \\
                                        H$\beta$ &  2.21 $\pm$  0.42 &  27.53 $\pm$  0.49 &   9.42 $\pm$  0.51 &  24.73 $\pm$  0.54 &  20.15 $\pm$  0.67 & 24.24 $\pm$ 1.00 &  29.14 $\pm$  0.62 &   \nodata                 &   \nodata                 &   \nodata                 \\
                                       H$\alpha$ & 13.89 $\pm$  1.08 &114.01 $\pm$ 1.44 & 45.09 $\pm$  1.54 & 141.08 $\pm$ 1.90 & 84.21 $\pm$  1.94 & 120.85 $\pm$ 2.24 &126.01 $\pm$ 1.96 &  \nodata                 &   \nodata                 &   \nodata                 \\
             \ion{He}{1} $\lambda$4473 &  $<$1.50               &   0.37 $\pm$  0.04 &   $<$1.61              &   0.43 $\pm$  0.08 &   1.48 $\pm$  0.45 &   $<$2.36               &   1.77 $\pm$  0.38 &    \nodata                 &   \nodata                 &   \nodata                 \\
             \ion{He}{2} $\lambda$4687 &  $<$1.74              &   3.87 $\pm$  0.35 &   2.85 $\pm$  0.42 &   1.58 $\pm$  0.43 &   6.15 $\pm$  0.44 &   7.94 $\pm$  0.42 &   2.65 $\pm$  0.30 &    \nodata                 &   \nodata                 &   \nodata                 \\
             \ion{He}{1} $\lambda$5877 &  2.18 $\pm$  0.45 &   7.83 $\pm$  0.59 &   4.97 $\pm$  0.55 &   5.48 $\pm$  0.62 & 14.89 $\pm$  0.64 & 21.48 $\pm$  0.84 &  13.20 $\pm$  0.54 &   \nodata                 &   \nodata                 &   \nodata                 \\
             \ion{He}{1} $\lambda$6679 &  $<$2.86               &   7.43 $\pm$  0.86 &   3.55 $\pm$  0.82 &   3.27 $\pm$  0.89 &   8.45 $\pm$  0.91 &    \nodata               &   5.89 $\pm$  0.73 &    \nodata                 &   \nodata                 &   \nodata                 \\
             \ion{He}{1} $\lambda$7067 &  $<$3.01               &    \nodata               &   4.81 $\pm$  0.97 &   6.02 $\pm$  1.06 & 20.57 $\pm$  1.10 &    \nodata               &  17.06 $\pm$  0.90 &   \nodata                 &   \nodata                 &   \nodata                 \\
    \ion{Fe}{2} $\lambda$4434-4684 &  $<$0.32               &   2.54 $\pm$  0.52 &   $<$0.18               &   $<$0.99               &   2.21 $\pm$  0.49 &   $<$2.29               &   3.51 $\pm$  0.51 &    \nodata                 &   \nodata                 &   \nodata                 \\
\hline\multicolumn{11}{c}{Power-law continuum}\\\hline
                                       $F_{5100}$ &  5.10 $\pm$  0.13 &  17.77 $\pm$  0.22 &  13.12 $\pm$  0.15 &  14.48 $\pm$  0.19 &   9.59 $\pm$  0.24 &  23.16 $\pm$  0.24 &  26.32 $\pm$  0.23 &   4.92 $\pm$  0.12 &   5.83 $\pm$  0.12 &   6.46 $\pm$  0.12 \\
                                             Slope & $-$1.64 $\pm$ 0.20 & $-$0.05 $\pm$ 0.06 & $-$0.54 $\pm$ 0.08 & 0.03 $\pm$ 0.07 & $-$0.68 $\pm$ 0.12 & $-$0.31 $\pm$ 0.05 & $-$0.46 $\pm$ 0.04 & $-$2.09 $\pm$ 0.17 & $-$1.41 $\pm$ 0.16 & $-$2.07 $\pm$ 0.15 \\
\enddata
\tablecomments{Line fluxes and continuum flux densities at 5100 \AA\ ($F_{5100}$) are given in units of $10^{-18}$ erg s$^{-1}$ cm$^{-2}$ and $10^{-21}$ erg s$^{-1}$ cm$^{-2}$ \AA$^{-1}$, respectively. The flux upper bounds are defined as the 3$\sigma$ upper limits.}
\end{deluxetable*}
\end{longrotatetable}

\begin{longrotatetable}
\begin{deluxetable*}{ccccccccccccccccccc}
\tablecaption{Emission line fluxes and continuum properties of G11, Q01, and Q02 \label{tab:line_fluxes2}}
\tablewidth{700pt}
\tabletypesize{\scriptsize}
\tablehead{
\colhead{} &                \colhead{G11} &                 \colhead{Q01} &                 \colhead{Q02}
} 
\startdata
\hline\multicolumn{4}{c}{SF/NLR lines}\\\hline
                                                   H9 &  \nodata               &   \nodata               &   \nodata               \\
                                                   H8 &  \nodata               &   \nodata               &   \nodata               \\
                                     H$\epsilon$ &  \nodata               &   \nodata               &   2.70 $\pm$  0.07 \\
                                         H$\delta$ &  2.09 $\pm$  0.15 &   7.30 $\pm$  0.18 &   2.99 $\pm$  0.08 \\
                                     H$\gamma$ &  4.10 $\pm$  0.19 &  12.32 $\pm$  0.24 &   5.51 $\pm$  0.09 \\
                                          H$\beta$ &  9.07 $\pm$  0.30 &  33.34 $\pm$  0.35 &  14.76 $\pm$  0.14 \\
                                        H$\alpha$ & 32.99 $\pm$  0.99 & 243.83 $\pm$  1.64 & 110.49 $\pm$  0.71 \\
             \ion{He}{1} $\lambda$4473 &  0.77 $\pm$  0.09 &   2.18 $\pm$  0.12 &   0.91 $\pm$  0.05 \\
             \ion{He}{2} $\lambda$4687 &  $<$0.40               &   1.07 $\pm$  0.14 &   1.49 $\pm$  0.06 \\
             \ion{He}{1} $\lambda$5877 &  1.48 $\pm$  0.12 &  11.05 $\pm$  0.21 &   6.73 $\pm$  0.09 \\
             \ion{He}{1} $\lambda$6679 &  0.84 $\pm$  0.13 &   4.24 $\pm$  0.23 &   3.06 $\pm$  0.10 \\
             \ion{He}{1} $\lambda$7067 &  0.96 $\pm$  0.15 &  15.67 $\pm$  0.30 &   9.03 $\pm$  0.13 \\
   $[$\ion{Ne}{3}$]$ $\lambda$3869 &  \nodata               &   \nodata               &   \nodata               \\
    $[$\ion{O}{3}$]$ $\lambda$4363 &  0.87 $\pm$  0.12 &  20.40 $\pm$  0.33 &   6.61 $\pm$  0.11 \\
    $[$\ion{O}{3}$]$ $\lambda$5007 & 58.35 $\pm$  1.17 & 157.77 $\pm$  0.89 &  45.36 $\pm$  0.29 \\
    $[$\ion{O}{1}$]$ $\lambda$6300 &  0.39 $\pm$  0.09 &   2.27 $\pm$  0.18 &   1.40 $\pm$  0.08 \\
    $[$\ion{N}{2}$]$ $\lambda$6585 &  5.10 $\pm$  0.17 &   $<$0.92               &   $<$0.05        \\
    $[$\ion{S}{2}$]$ $\lambda$6716 &  0.81 $\pm$  0.11 &   0.98 $\pm$  0.22 &   0.44 $\pm$  0.10 \\
    $[$\ion{S}{2}$]$ $\lambda$6731 &  0.86 $\pm$  0.13 &   0.72 $\pm$  0.19 &   0.43 $\pm$  0.09 \\
\hline\multicolumn{4}{c}{BLR lines}\\\hline
                                                  H9 &  \nodata               &   \nodata               &   \nodata               \\
                                                  H8 &  \nodata               &   \nodata               &   \nodata               \\
                                    H$\epsilon$ &  \nodata               &   \nodata               &   3.47 $\pm$  0.09 \\
                                        H$\delta$ &  \nodata               &   6.65 $\pm$  0.19 &   5.61 $\pm$  0.34 \\
                                    H$\gamma$ &  \nodata               &  23.80 $\pm$  0.90 &  17.97 $\pm$  0.36 \\
                                         H$\beta$ &  \nodata               &  92.88 $\pm$  1.01 &  63.52 $\pm$  0.39 \\
                                       H$\alpha$ &  \nodata               & 541.20 $\pm$  4.73 & 391.17 $\pm$  1.41 \\
             \ion{He}{1} $\lambda$4473 &  \nodata               &   1.99 $\pm$  0.11 &   1.17 $\pm$  0.06 \\
             \ion{He}{2} $\lambda$4687 &  \nodata               &  10.69 $\pm$  0.58 &  10.23 $\pm$  0.27 \\
             \ion{He}{1} $\lambda$5877 &  \nodata               &  31.44 $\pm$  0.69 &  24.19 $\pm$  0.34 \\
             \ion{He}{1} $\lambda$6679 &  \nodata               &   9.11 $\pm$  1.37 &  10.18 $\pm$  0.63 \\
             \ion{He}{1} $\lambda$7067 &  \nodata               &  31.74 $\pm$  1.01 &  33.08 $\pm$  0.55 \\
    \ion{Fe}{2} $\lambda$4434-4684 &  \nodata               &   5.83 $\pm$  0.54 &   5.35 $\pm$  0.24 \\
\hline\multicolumn{4}{c}{Power-law continuum}\\\hline
                                     $F_{5100}$ & 17.91 $\pm$  0.07 &  73.00 $\pm$  0.23 &  35.10 $\pm$  0.10 \\
                                              Slope & $-$2.10 $\pm$  0.03 &  $-$0.47 $\pm$  0.01 &  $-$0.39 $\pm$  0.01 \\
\enddata
\tablecomments{The same notes apply as in Table \ref{tab:line_fluxes1}.}
\end{deluxetable*}
\end{longrotatetable}


\begin{deluxetable*}{ccccc}
\tablecaption{Alternative estimates of $L_{\rm bol}$ and $\lambda_{\rm Edd}$ \label{tab:alternative_Lbol}}
\tablewidth{0pt}
\tablehead{
\colhead{} & \multicolumn{2}{c}{$L_{\rm 5100}$-based} &  \multicolumn{2}{c}{$L_{\rm [O III]}$-based} \\
\colhead{Name} & \colhead{$\log L_{\rm bol}$} & \colhead{$\lambda_{\rm Edd}$} &  \colhead{$\log L_{\rm bol}$} & \colhead{$\lambda_{\rm Edd}$} 
}
\startdata
G01 & 45.81 $\pm$ 0.25 &   0.81 $\pm$ 0.50 &  46.42 $\pm$ 0.02 &   3.33 $\pm$ 0.75\\
G02 & 45.90 $\pm$ 0.03 &   0.32 $\pm$ 0.02 &  46.64 $\pm$ 0.03 &   1.75 $\pm$ 0.13\\
G03 & 45.88 $\pm$ 0.08 &   0.37 $\pm$ 0.07 &  46.77 $\pm$ 0.02 &   2.92 $\pm$ 0.25\\
G04 & 46.14 $\pm$ 0.03 &   0.39 $\pm$ 0.03 &  46.78 $\pm$ 0.04 &   1.66 $\pm$ 0.15\\
G05 & 45.59 $\pm$ 0.05 &   0.32 $\pm$ 0.04 &  46.18 $\pm$ 0.05 &   1.27 $\pm$ 0.14\\
G06 & 46.34 $\pm$ 0.05 &   0.46 $\pm$ 0.06 &  47.18 $\pm$ 0.06 &   3.15 $\pm$ 0.44\\
G07 & 46.08 $\pm$ 0.03 &   0.87 $\pm$ 0.07 &  46.66 $\pm$ 0.04 &   3.31 $\pm$ 0.31\\
G08 & 44.87 $\pm$ 0.09 &   \nodata               &  45.71 $\pm$ 0.09 &       \nodata\\
G09 & 45.06 $\pm$ 0.09 &   \nodata               &  46.29 $\pm$ 0.10 &       \nodata\\
G10 & 44.99 $\pm$ 0.04 &   \nodata               &  46.07 $\pm$ 0.05 &       \nodata\\
G11 & 45.71 $\pm$ 0.04 &   \nodata               &  46.48 $\pm$ 0.04 &       \nodata\\
Q01 & 46.88 $\pm$ 0.02 &   0.47 $\pm$ 0.02 &  47.77 $\pm$ 0.01 &   3.60 $\pm$ 0.13\\
Q02 & 46.66 $\pm$ 0.01 &   0.34 $\pm$  0.01 &  47.27 $\pm$ 0.01 &   1.40 $\pm$ 0.04\\
\enddata
\tablecomments{
The bolometric luminosities ($L_{\rm bol}$) are presented in units of erg s$^{-1}$.
}
\end{deluxetable*}


\begin{thebibliography}{}


\bibitem[Aihara et al.(2022)]{aihara22} Aihara, H., AlSayyad, Y., Ando, M., et al.\ 2022, \pasj, 74, 247. doi:10.1093/pasj/psab122
\bibitem[Aihara et al.(2018)]{aihara18} Aihara, H., Arimoto, N., Armstrong, R., et al.\ 2018, \pasj, 70, S4. doi:10.1093/pasj/psx066
\bibitem[Akins et al.(2025)]{akins25} Akins, H.~B., Casey, C.~M., Berg, D.~A., et al.\ 2025, \apjl, 980, 2, L29. doi:10.3847/2041-8213/adab76
\bibitem[Akins et al.(2024)]{akins24} Akins, H.~B., Casey, C.~M., Lambrides, E., et al.\ 2024, arXiv:2406.10341. doi:10.48550/arXiv.2406.10341
\bibitem[Alexandroff et al.(2013)]{alexandroff13} Alexandroff, R., Strauss, M.~A., Greene, J.~E., et al.\ 2013, \mnras, 435, 4, 3306. doi:10.1093/mnras/stt1500
\bibitem[Ananna et al.(2024)]{ananna24} Ananna, T.~T., Bogd{\'a}n, {\'A}., Kov{\'a}cs, O.~E., et al.\ 2024, \apjl, 969, L18. doi:10.3847/2041-8213/ad5669
\bibitem[Baldwin et al.(1981)]{baldwin81} Baldwin, J.~A., Phillips, M.~M., \& Terlevich, R.\ 1981, \pasp, 93, 5. doi:10.1086/130766
\bibitem[Baggen et al.(2024)]{baggen24} Baggen, J.~F.~W., van Dokkum, P., Brammer, G., et al.\ 2024, \apjl, 977, L13. doi:10.3847/2041-8213/ad90b8
\bibitem[Barro et al.(2024)]{barro24} Barro, G., P{\'e}rez-Gonz{\'a}lez, P.~G., Kocevski, D.~D., et al.\ 2024, \apj, 963, 128. doi:10.3847/1538-4357/ad167e
\bibitem[Boroson \& Green(1992)]{boroson92} Boroson, T.~A. \& Green, R.~F.\ 1992, \apjs, 80, 109. doi:10.1086/191661
\bibitem[Bogd{\'a}n et al.(2024)]{bogdan24} Bogd{\'a}n, {\'A}., Goulding, A.~D., Natarajan, P., et al.\ 2024, Nature Astronomy, 8, 126. doi:10.1038/s41550-023-02111-9
\bibitem[Bouwens et al.(2021)]{bouwens21} Bouwens, R.~J., Oesch, P.~A., Stefanon, M., et al.\ 2021, \aj, 162, 47. doi:10.3847/1538-3881/abf83e
\bibitem[Bowler et al.(2015)]{bowler15} Bowler, R.~A.~A., Dunlop, J.~S., McLure, R.~J., et al.\ 2015, \mnras, 452, 1817. doi:10.1093/mnras/stv1403
\bibitem[Boyett et al.(2024)]{boyett24} Boyett, K., Bunker, A.~J., Curtis-Lake, E., et al.\ 2024, \mnras, 535, 2, 1796. doi:10.1093/mnras/stae2430
\bibitem[Calzetti et al.(2000)]{calzetti00} Calzetti, D., Armus, L., Bohlin, R.~C., et al.\ 2000, \apj, 533, 682. doi:10.1086/308692
\bibitem[Castellano et al.(2022)]{castellano22} Castellano, M., Fontana, A., Treu, T., et al.\ 2022, \apjl, 938, L15. doi:10.3847/2041-8213/ac94d0
\bibitem[Carnall et al.(2023)]{carnall23} Carnall, A.~C., McLure, R.~J., Dunlop, J.~S., et al.\ 2023, \nat, 619, 716. doi:10.1038/s41586-023-06158-6
\bibitem[Carniani et al.(2024a)]{carniani24a} Carniani, S., D'Eugenio, F., Ji, X., et al.\ 2024, arXiv:2409.20533. doi:10.48550/arXiv.2409.20533
\bibitem[Carniani et al.(2024b)]{carniani24b} Carniani, S., Hainline, K., D'Eugenio, F., et al.\ 2024, \nat, 633, 318. doi:10.1038/s41586-024-07860-9
\bibitem[Casey et al.(2024)]{casey24} Casey, C.~M., Akins, H.~B., Kokorev, V., et al.\ 2024, \apjl, 975, L4. doi:10.3847/2041-8213/ad7ba7
\bibitem[Clarke et al.(2024)]{clarke24} Clarke, L., Shapley, A.~E., Sanders, R.~L., et al.\ 2024, \apj, 977, 133. doi:10.3847/1538-4357/ad8ba4
\bibitem[Decarli et al.(2017)]{decarli17} Decarli, R., Walter, F., Venemans, B.~P., et al.\ 2017, \nat, 545, 457. doi:10.1038/nature22358
\bibitem[Decarli et al.(2019a)]{decarli19a} Decarli, R., Dotti, M., Ba{\~n}ados, E., et al.\ 2019, \apj, 880, 157. doi:10.3847/1538-4357/ab297f
\bibitem[Decarli et al.(2019b)]{decarli19b} Decarli, R., Mignoli, M., Gilli, R., et al.\ 2019, \aap, 631, L10. doi:10.1051/0004-6361/201936813
\bibitem[de Graaff et al.(2024)]{degraaff24} de Graaff, A., Rix, H.-W., Carniani, S., et al.\ 2024, \aap, 684, A87. doi:10.1051/0004-6361/202347755
\bibitem[Ding et al.(2025)]{ding25} Ding, X., Onoue, M., Silverman, J.~D., et al.\ 2025, arXiv:2505.03876. doi:10.48550/arXiv.2505.03876
\bibitem[Ding et al.(2023)]{ding23} Ding, X., Onoue, M., Silverman, J.~D., et al.\ 2023, \nat, 621, 51. doi:10.1038/s41586-023-06345-5
\bibitem[Dom{\'\i}nguez et al.(2013)]{dominguez13} Dom{\'\i}nguez, A., Siana, B., Henry, A.~L., et al.\ 2013, \apj, 763, 145. doi:10.1088/0004-637X/763/2/145
\bibitem[Donnan et al.(2024)]{donnan24} Donnan, C.~T., McLure, R.~J., Dunlop, J.~S., et al.\ 2024, \mnras, 533, 3222. doi:10.1093/mnras/stae2037
\bibitem[Fan et al.(2023)]{fan23} Fan, X., Ba{\~n}ados, E., \& Simcoe, R.~A.\ 2023, \araa, 61, 373. doi:10.1146/annurev-astro-052920-102455
\bibitem[Feltre et al.(2016)]{feltre16} Feltre, A., Charlot, S., \& Gutkin, J.\ 2016, \mnras, 456, 3354. doi:10.1093/mnras/stv2794
\bibitem[Finkelstein et al.(2023)]{finkelstein23} Finkelstein, S.~L., Bagley, M.~B., Ferguson, H.~C., et al.\ 2023, \apjl, 946, L13. doi:10.3847/2041-8213/acade4
\bibitem[Finkelstein et al.(2015)]{finkelstein15} Finkelstein, S.~L., Ryan, R.~E., Papovich, C., et al.\ 2015, \apj, 810, 71. doi:10.1088/0004-637X/810/1/71
\bibitem[Freeman et al.(2019)]{freeman19} Freeman, W.~R., Siana, B., Kriek, M., et al.\ 2019, \apj, 873, 102. doi:10.3847/1538-4357/ab0655
\bibitem[Fujimoto et al.(2024)]{fujimoto24} Fujimoto, S., Wang, B., Weaver, J.~R., et al.\ 2024, \apj, 977, 250. doi:10.3847/1538-4357/ad9027
\bibitem[Furtak et al.(2024)]{furtak24} Furtak, L.~J., Labb{\'e}, I., Zitrin, A., et al.\ 2024, \nat, 628, 57. doi:10.1038/s41586-024-07184-8
\bibitem[Genzel et al.(2011)]{genzel11} Genzel, R., Newman, S., Jones, T., et al.\ 2011, \apj, 733, 101. doi:10.1088/0004-637X/733/2/101
\bibitem[Greene et al.(2014)]{greene14} Greene, J.~E., Alexandroff, R., Strauss, M.~A., et al.\ 2014, \apj, 788, 1, 91. doi:10.1088/0004-637X/788/1/91
\bibitem[Greene \& Ho(2005)]{greene05} Greene, J.~E. \& Ho, L.~C.\ 2005, \apj, 630, 122. doi:10.1086/431897
\bibitem[Greene et al.(2024)]{greene24} Greene, J.~E., Labbe, I., Goulding, A.~D., et al.\ 2024, \apj, 964, 39. doi:10.3847/1538-4357/ad1e5f
\bibitem[Gloudemans et al.(2025)]{gloudemans25} Gloudemans, A.~J., Duncan, K.~J., Eilers, A.-C., et al.\ 2025, arXiv:2501.04912. doi:10.48550/arXiv.2501.04912
\bibitem[Goulding et al.(2023)]{goulding23} Goulding, A.~D., Greene, J.~E., Setton, D.~J., et al.\ 2023, \apjl, 955, L24. doi:10.3847/2041-8213/acf7c5
\bibitem[Gutkin et al.(2016)]{gutkin16} Gutkin, J., Charlot, S., \& Bruzual, G.\ 2016, \mnras, 462, 1757. doi:10.1093/mnras/stw1716
\bibitem[Harikane et al.(2022)]{harikane22} Harikane, Y., Ono, Y., Ouchi, M., et al.\ 2022, \apjs, 259, 20. doi:10.3847/1538-4365/ac3dfc
\bibitem[Harikane et al.(2023)]{harikane23} Harikane, Y., Zhang, Y., Nakajima, K., et al.\ 2023, \apj, 959, 39. doi:10.3847/1538-4357/ad029e
\bibitem[Heckman \& Best(2014)]{heckman14} Heckman, T.~M. \& Best, P.~N.\ 2014, \araa, 52, 589. doi:10.1146/annurev-astro-081913-035722
\bibitem[Inayoshi et al.(2024)]{inayoshi24} Inayoshi, K., Kimura, S., \& Noda, H.\ 2024, arXiv:2412.03653. doi:10.48550/arXiv.2412.03653
\bibitem[Inoue et al.(2020)]{inoue20} Inoue, A.~K., Yamanaka, S., Ouchi, M., et al.\ 2020, \pasj, 72, 101. doi:10.1093/pasj/psaa100
\bibitem[Iwamoto et al.(2025)]{iwamoto25} Iwamoto, R., Matsuoka, Y., Imanishi, M., et al.\ 2025, \apj, 979, 183. doi:10.3847/1538-4357/ada27b
\bibitem[Iwasawa et al.(2025)]{iwasawa25} Iwasawa, K., Gilli, R., Vito, F., et al.\ 2025, arXiv:2505.04826. doi:10.48550/arXiv.2505.04826
\bibitem[Izumi et al.(2021a)]{izumi21a} Izumi, T., Matsuoka, Y., Fujimoto, S., et al.\ 2021, \apj, 914, 36. doi:10.3847/1538-4357/abf6dc
\bibitem[Izumi et al.(2021b)]{izumi21b} Izumi, T., Onoue, M., Matsuoka, Y., et al.\ 2021, \apj, 908, 235. doi:10.3847/1538-4357/abd7ef
\bibitem[Izumi et al.(2019)]{izumi19} Izumi, T., Onoue, M., Matsuoka, Y., et al.\ 2019, \pasj, 71, 111. doi:10.1093/pasj/psz096
\bibitem[Izumi et al.(2018)]{izumi18} Izumi, T., Onoue, M., Shirakata, H., et al.\ 2018, \pasj, 70, 36. doi:10.1093/pasj/psy026
\bibitem[Juod{\v{z}}balis et al.(2024)]{juodzbalis24} Juod{\v{z}}balis, I., Maiolino, R., Baker, W.~M., et al.\ 2024, \nat, 636, 594. doi:10.1038/s41586-024-08210-5
\bibitem[Juod{\v{z}}balis et al.(2025)]{juodzbalis25} Juod{\v{z}}balis, I., Maiolino, R., Baker, W.~M., et al.\ 2025, arXiv:2504.03551. doi:10.48550/arXiv.2504.03551
\bibitem[Kauffmann \& Heckman(2009)]{kauffmann09} Kauffmann, G. \& Heckman, T.~M.\ 2009, \mnras, 397, 135. doi:10.1111/j.1365-2966.2009.14960.x
\bibitem[Kauffmann et al.(2003)]{kauffmann03} Kauffmann, G., Heckman, T.~M., Tremonti, C., et al.\ 2003, \mnras, 346, 1055. doi:10.1111/j.1365-2966.2003.07154.x
\bibitem[Kawanomoto et al.(2018)]{kawanomoto18} Kawanomoto, S., Uraguchi, F., Komiyama, Y., et al.\ 2018, \pasj, 70, 66. doi:10.1093/pasj/psy056
\bibitem[Kato et al.(2020)]{kato20} Kato, N., Matsuoka, Y., Onoue, M., et al.\ 2020, \pasj, 72, 84. doi:10.1093/pasj/psaa074
\bibitem[Kennicutt(1998)]{kennicutt98} Kennicutt, R.~C.\ 1998, \araa, 36, 189. doi:10.1146/annurev.astro.36.1.189
\bibitem[Kewley et al.(2001)]{kewley01} Kewley, L.~J., Dopita, M.~A., Sutherland, R.~S., et al.\ 2001, \apj, 556, 121. doi:10.1086/321545
\bibitem[Killi et al.(2024)]{killi24} Killi, M., Watson, D., Brammer, G., et al.\ 2024, \aap, 691, A52. doi:10.1051/0004-6361/202348857
\bibitem[Kocevski et al.(2024)]{kocevski24} Kocevski, D.~D., Finkelstein, S.~L., Barro, G., et al.\ 2024, arXiv:2404.03576. doi:10.48550/arXiv.2404.03576
\bibitem[Kocevski et al.(2023)]{kocevski23} Kocevski, D.~D., Onoue, M., Inayoshi, K., et al.\ 2023, \apjl, 954, L4. doi:10.3847/2041-8213/ace5a0
\bibitem[Kokorev et al.(2023)]{kokorev23} Kokorev, V., Fujimoto, S., Labbe, I., et al.\ 2023, \apjl, 957, L7. doi:10.3847/2041-8213/ad037a
\bibitem[Kokubo \& Harikane(2024)]{kokubo24} Kokubo, M. \& Harikane, Y.\ 2024, arXiv:2407.04777. doi:10.48550/arXiv.2407.04777
\bibitem[Konno et al.(2016)]{konno16} Konno, A., Ouchi, M., Nakajima, K., et al.\ 2016, \apj, 823, 20 
\bibitem[Korista \& Goad(2004)]{korista04} Korista, K.~T. \& Goad, M.~R.\ 2004, \apj, 606, 749. doi:10.1086/383193
\bibitem[Kulas et al.(2012)]{kulas12} Kulas, K.~R., Shapley, A.~E., Kollmeier, J.~A., et al.\ 2012, \apj, 745, 33. doi:10.1088/0004-637X/745/1/33
\bibitem[Labb{\'e} et al.(2023)]{labbe23} Labb{\'e}, I., van Dokkum, P., Nelson, E., et al.\ 2023, \nat, 616, 266. doi:10.1038/s41586-023-05786-2
\bibitem[Labb{\'e} et al.(2025)]{labbe25} Labbe, I., Greene, J.~E., Bezanson, R., et al.\ 2025, \apj, 978, 92. doi:10.3847/1538-4357/ad3551
\bibitem[Labb{\'e} et al.(2024)]{labbe24} Labbe, I., Greene, J.~E., Matthee, J., et al.\ 2024, arXiv:2412.04557. doi:10.48550/arXiv.2412.04557
\bibitem[Larson et al.(2023)]{larson23} Larson, R.~L., Finkelstein, S.~L., Kocevski, D.~D., et al.\ 2023, \apjl, 953, L29. doi:10.3847/2041-8213/ace619
\bibitem[Liddle(2007)]{liddle07} Liddle, A.~R.\ 2007, \mnras, 377, L74. doi:10.1111/j.1745-3933.2007.00306.x
\bibitem[Lin et al.(2025)]{lin25} Lin, X., Fan, X., Wang, F., et al.\ 2025, arXiv:2504.08039. doi:10.48550/arXiv.2504.08039
\bibitem[Lin et al.(2024)]{lin24} Lin, X., Wang, F., Fan, X., et al.\ 2024, \apj, 974, 147. doi:10.3847/1538-4357/ad6565
\bibitem[Llerena et al.(2024)]{llerena24} Llerena, M., Amor{\'\i}n, R., Pentericci, L., et al.\ 2024, \aap, 691, A59. doi:10.1051/0004-6361/202449904
\bibitem[Lyu et al.(2024)]{lyu24} Lyu, J., Rieke, G.~H., Stone, M., et al.\ 2024, arXiv:2412.04548. doi:10.48550/arXiv.2412.04548
\bibitem[Ma et al.(2025a)]{ma25} Ma, Y., Greene, J.~E., Setton, D.~J., et al.\ 2025, \apj, 981, 191. doi:10.3847/1538-4357/ada613
\bibitem[Ma et al.(2025b)]{ma25b} Ma, Y., Greene, J.~E., Setton, D.~J., et al.\ 2025, arXiv:2504.08032. doi:10.48550/arXiv.2504.08032
\bibitem[Maiolino et al.(2024a)]{maiolino24_xray} Maiolino, R., Risaliti, G., Signorini, M., et al.\ 2024, arXiv:2405.00504. doi:10.48550/arXiv.2405.00504
\bibitem[Maiolino et al.(2024b)]{maiolino24} Maiolino, R., Scholtz, J., Curtis-Lake, E., et al.\ 2024, \aap, 691, A145. doi:10.1051/0004-6361/202347640
\bibitem[Matsuoka et al.(2018a)]{p4} Matsuoka, Y., Iwasawa, K., Onoue, M., et al.\ 2018a, \apjs, 237, 5
\bibitem[Matsuoka et al.(2019a)]{p10} Matsuoka, Y., Iwasawa, K., Onoue, M., et al.\ 2019, \apj, 883, 183. doi:10.3847/1538-4357/ab3c60
\bibitem[Matsuoka et al.(2022)]{p16} Matsuoka, Y., Iwasawa, K., Onoue, M., et al.\ 2022, \apjs, 259, 18. doi:10.3847/1538-4365/ac3d31
\bibitem[Matsuoka et al.(2024)]{p20} Matsuoka, Y., Izumi, T., Onoue, M., et al.\ 2024, \apjl, 965, L4. doi:10.3847/2041-8213/ad35c7
\bibitem[Matsuoka et al.(2008)]{matsuoka08} Matsuoka, Y., Kawara, K., \& Oyabu, S.\ 2008, \apj, 673, 62. doi:10.1086/524193
\bibitem[Matsuoka et al.(2023)]{p19} Matsuoka, Y., Onoue, M., Iwasawa, K., et al.\ 2023, \apjl, 949, L42. doi:10.3847/2041-8213/acd69f
\bibitem[Matsuoka et al.(2016)]{p1} Matsuoka, Y., Onoue, M., Kashikawa, N., et al.\ 2016, \apj, 828, 26
\bibitem[Matsuoka et al.(2018b)]{p2} Matsuoka, Y., Onoue, M., Kashikawa, N., et al.\ 2018b, \pasj, 70, S35
\bibitem[Matsuoka et al.(2019b)]{p7} Matsuoka, Y., Onoue, M., Kashikawa, N., et al.\ 2019, \apjl, 872, L2. doi:10.3847/2041-8213/ab0216
\bibitem[Matsuoka et al.(2007)]{matsuoka07} Matsuoka, Y., Oyabu, S., Tsuzuki, Y., et al.\ 2007, \apj, 663, 781. doi:10.1086/518399
\bibitem[Matsuoka et al.(2018c)]{p5} Matsuoka, Y., Strauss, M.~A., Kashikawa, N., et al.\ 2018c, \apj, 869, 150 
\bibitem[Matthee et al.(2024)]{matthee24} Matthee, J., Naidu, R.~P., Brammer, G., et al.\ 2024, \apj, 963, 129. doi:10.3847/1538-4357/ad2345
\bibitem[Matthee et al.(2018)]{matthee18} Matthee, J., Sobral, D., Gronke, M., et al.\ 2018, \aap, 619, A136. doi:10.1051/0004-6361/201833528
\bibitem[Mazzolari et al.(2024a)]{mazzolari24} Mazzolari, G., Gilli, R., Maiolino, R., et al.\ 2024, arXiv:2412.04224. doi:10.48550/arXiv.2412.04224
\bibitem[Mazzolari et al.(2024b)]{mazzolari24b} Mazzolari, G., Scholtz, J., Maiolino, R., et al.\ 2024, arXiv:2408.15615. doi:10.48550/arXiv.2408.15615
\bibitem[Mazzolari et al.(2024c)]{mazzolari24c} Mazzolari, G., {\"U}bler, H., Maiolino, R., et al.\ 2024, \aap, 691, A345. doi:10.1051/0004-6361/202450407
\bibitem[Miyazaki et al.(2018)]{miyazaki18} Miyazaki, S., Komiyama, Y., Kawanomoto, S., et al.\ 2018, \pasj, 70, S1. doi:10.1093/pasj/psx063
\bibitem[Naidu et al.(2025a)]{naidu25_lrd} Naidu, R.~P., Matthee, J., Katz, H., et al.\ 2025, arXiv:2503.16596. doi:10.48550/arXiv.2503.16596
\bibitem[Naidu et al.(2025b)]{naidu25_gal} Naidu, R.~P., Oesch, P.~A., Brammer, G., et al.\ 2025, arXiv:2505.11263. 
\bibitem[Nakajima \& Maiolino(2022)]{nakajima22} Nakajima, K. \& Maiolino, R.\ 2022, \mnras, 513, 5134. doi:10.1093/mnras/stac1242
\bibitem[Newman et al.(2012)]{newman12} Newman, S.~F., Genzel, R., F{\"o}rster-Schreiber, N.~M., et al.\ 2012, \apj, 761, 43. doi:10.1088/0004-637X/761/1/43
\bibitem[Oke \& Gunn(1983)]{oke83} Oke, J.~B. \& Gunn, J.~E.\ 1983, \apj, 266, 713. doi:10.1086/160817
\bibitem[Onoue et al.(2024)]{onoue24} Onoue, M., Ding, X., Silverman, J.~D., et al.\ 2024, arXiv:2409.07113. doi:10.48550/arXiv.2409.07113
\bibitem[Onoue et al.(2019)]{onoue19} Onoue, M., Kashikawa, N., Matsuoka, Y., et al.\ 2019, \apj, 880, 77. doi:10.3847/1538-4357/ab29e9
\bibitem[Onoue et al.(2021)]{onoue21} Onoue, M., Matsuoka, Y., Kashikawa, N., et al.\ 2021, \apj, 919, 61. doi:10.3847/1538-4357/ac0f07
\bibitem[Onoue et al.(2023)]{onoue23} Onoue, M., Inayoshi, K., Ding, X., et al.\ 2023, \apjl, 942, L17. doi:10.3847/2041-8213/aca9d3
\bibitem[Osterbrock \& Ferland(2006)]{osterbrock06} Osterbrock, D.~E. \& Ferland, G.~J.\ 2006, , Astrophysics of gaseous nebulae and active galactic nuclei. 
\bibitem[Ouchi et al.(2018)]{ouchi18} Ouchi, M., Harikane, Y., Shibuya, T., et al.\ 2018, \pasj, 70, S13. doi:10.1093/pasj/psx074
\bibitem[Pacucci \& Narayan(2024)]{pacucci24} Pacucci, F. \& Narayan, R.\ 2024, \apj, 976, 96. doi:10.3847/1538-4357/ad84f7
\bibitem[Perger et al.(2025)]{perger25} Perger, K., Fogasy, J., Frey, S., et al.\ 2025, \aap, 693, L2. doi:10.1051/0004-6361/202452422
\bibitem[Pei(1992)]{pei92} Pei, Y.~C.\ 1992, \apj, 395, 130. doi:10.1086/171637
\bibitem[Rakshit et al.(2020)]{rakshit20} Rakshit, S., Stalin, C.~S., \& Kotilainen, J.\ 2020, \apjs, 249, 17. doi:10.3847/1538-4365/ab99c5
\bibitem[Reines et al.(2013)]{reines13} Reines, A.~E., Greene, J.~E., \& Geha, M.\ 2013, \apj, 775, 116. doi:10.1088/0004-637X/775/2/116
\bibitem[Rigby et al.(2023)]{rigby23} Rigby, J., Perrin, M., McElwain, M., et al.\ 2023, \pasp, 135, 1046, 048001. doi:10.1088/1538-3873/acb293
\bibitem[Robertson et al.(2024)]{robertson24} Robertson, B., Johnson, B.~D., Tacchella, S., et al.\ 2024, \apj, 970, 31. doi:10.3847/1538-4357/ad463d
\bibitem[Sawamura et al.(2025)]{sawamura25} Sawamura, M., Izumi, T., Nakanishi, K., et al.\ 2025, \apj, 980, 121. doi:10.3847/1538-4357/ada943
\bibitem[Scholtz et al.(2023)]{scholtz23} Scholtz, J., Maiolino, R., D'Eugenio, F., et al.\ 2023, arXiv:2311.18731. doi:10.48550/arXiv.2311.18731
\bibitem[Schindler et al.(2024)]{schindler24} Schindler, J.-T., Hennawi, J.~F., Davies, F.~B., et al.\ 2024, arXiv:2411.11534. doi:10.48550/arXiv.2411.11534
\bibitem[Selsing et al.(2016)]{selsing16} Selsing, J., Fynbo, J.~P.~U., Christensen, L., et al.\ 2016, \aap, 585, A87. doi:10.1051/0004-6361/201527096
\bibitem[Setton et al.(2024)]{setton24} Setton, D.~J., Greene, J.~E., de Graaff, A., et al.\ 2024, arXiv:2411.03424. doi:10.48550/arXiv.2411.03424
\bibitem[Setton et al.(2025)]{setton25} Setton, D.~J., Greene, J.~E., Spilker, J.~S., et al.\ 2025, arXiv:2503.02059. doi:10.48550/arXiv.2503.02059
\bibitem[Shapiro et al.(2009)]{shapiro09} Shapiro, K.~L., Genzel, R., Quataert, E., et al.\ 2009, \apj, 701, 955. doi:10.1088/0004-637X/701/2/955
\bibitem[Shen et al.(2020)]{shen20} Shen, X., Hopkins, P.~F., Faucher-Gigu{\`e}re, C.-A., et al.\ 2020, \mnras, 495, 3252. doi:10.1093/mnras/staa1381
\bibitem[Shen et al.(2011)]{shen11} Shen, Y., Richards, G.~T., Strauss, M.~A., et al.\ 2011, \apjs, 194, 45. doi:10.1088/0067-0049/194/2/45
\bibitem[Shen et al.(2019)]{shen19} Shen, Y., Wu, J., Jiang, L., et al.\ 2019, \apj, 873, 1, 35. doi:10.3847/1538-4357/ab03d9
\bibitem[Shibuya et al.(2018)]{shibuya18} Shibuya, T., Ouchi, M., Konno, A., et al.\ 2018, \pasj, 70, S14. doi:10.1093/pasj/psx122
\bibitem[Sobral et al.(2018)]{sobral18} Sobral, D., Matthee, J., Darvish, B., et al.\ 2018, \mnras, 477, 2817. doi:10.1093/mnras/sty782
\bibitem[Songaila et al.(2018)]{songaila18} Songaila, A., Hu, E.~M., Barger, A.~J., et al.\ 2018, \apj, 859, 91. doi:10.3847/1538-4357/aac021
\bibitem[Spinoso et al.(2020)]{spinoso20} Spinoso, D., Orsi, A., L{\'o}pez-Sanjuan, C., et al.\ 2020, \aap, 643, A149. doi:10.1051/0004-6361/202038756
\bibitem[Stepney et al.(2024)]{stepney24} Stepney, M., Banerji, M., Tang, S., et al.\ 2024, \mnras, 533, 3, 2948. doi:10.1093/mnras/stae1970
\bibitem[Stern \& Laor(2012)]{stern12} Stern, J. \& Laor, A.\ 2012, \mnras, 423, 1, 600. doi:10.1111/j.1365-2966.2012.20901.x
\bibitem[Stone et al.(2024)]{stone24} Stone, M.~A., Lyu, J., Rieke, G.~H., et al.\ 2024, \apj, 964, 90. doi:10.3847/1538-4357/ad2a57
\bibitem[Storey \& Zeippen(2000)]{storey00} Storey, P.~J. \& Zeippen, C.~J.\ 2000, \mnras, 312, 4, 813. doi:10.1046/j.1365-8711.2000.03184.x
\bibitem[Swinbank et al.(2019)]{swinbank19} Swinbank, A.~M., Harrison, C.~M., Tiley, A.~L., et al.\ 2019, \mnras, 487, 381. doi:10.1093/mnras/stz1275
\bibitem[Takahashi et al.(2024)]{takahashi24} Takahashi, A., Matsuoka, Y., Onoue, M., et al.\ 2024, \apj, 960, 112. doi:10.3847/1538-4357/ad045e
\bibitem[Taylor et al.(2024)]{taylor24} Taylor, A.~J., Finkelstein, S.~L., Kocevski, D.~D., et al.\ 2024, arXiv:2409.06772. doi:10.48550/arXiv.2409.06772
\bibitem[Taylor et al.(2025)]{taylor25} Taylor, A.~J., Kokorev, V., Kocevski, D.~D., et al.\ 2025, arXiv:2505.04609. doi:10.48550/arXiv.2505.04609
\bibitem[Tee et al.(2024)]{tee24} Tee, W.~L., Fan, X., Wang, F., et al.\ 2024, arXiv:2412.05242. doi:10.48550/arXiv.2412.05242
\bibitem[Toba et al.(2021)]{toba21} Toba, Y., Ueda, Y., Gandhi, P., et al.\ 2021, \apj, 912, 2, 91. doi:10.3847/1538-4357/abe94a
\bibitem[Taylor et al.(2025)]{taylor25} Taylor, A.~J., Kokorev, V., Kocevski, D.~D., et al.\ 2025, arXiv:2505.04609. doi:10.48550/arXiv.2505.04609
\bibitem[Tripodi et al.(2024)]{tripodi24} Tripodi, R., Martis, N., Markov, V., et al.\ 2024, arXiv:2412.04983. doi:10.48550/arXiv.2412.04983
\bibitem[Tsuzuki et al.(2006)]{tsuzuki06} Tsuzuki, Y., Kawara, K., Yoshii, Y., et al.\ 2006, \apj, Fe II Emission in 14 Low-Redshift Quasars. I. Observations, 650, 1, 57. doi:10.1086/506376
\bibitem[{\"U}bler et al.(2023)]{ubler23} {\"U}bler, H., Maiolino, R., Curtis-Lake, E., et al.\ 2023, \aap, 677, A145. doi:10.1051/0004-6361/202346137
\bibitem[Umeda et al.(2024)]{umeda24} Umeda, H., Ouchi, M., Kikuta, S., et al.\ 2024, arXiv:2411.15495. doi:10.48550/arXiv.2411.15495
\bibitem[Vanden Berk et al.(2001)]{vandenberk01} Vanden Berk, D.~E., Richards, G.~T., Bauer, A., et al.\ 2001, \aj, 122, 549. doi:10.1086/321167
\bibitem[Verhamme et al.(2008)]{verhamme08} Verhamme, A., Schaerer, D., Atek, H., et al.\ 2008, \aap, 491, 89. doi:10.1051/0004-6361:200809648
\bibitem[Verhamme et al.(2006)]{verhamme06} Verhamme, A., Schaerer, D., \& Maselli, A.\ 2006, \aap, 460, 397. doi:10.1051/0004-6361:20065554
\bibitem[V{\'e}ron-Cetty et al.(2004)]{veron04} V{\'e}ron-Cetty, M.-P., Joly, M., \& V{\'e}ron, P.\ 2004, \aap, The unusual emission line spectrum of I Zw 1, 417, 515. doi:10.1051/0004-6361:20035714
\bibitem[Vestergaard \& Peterson(2006)]{vestergaard06} Vestergaard, M. \& Peterson, B.~M.\ 2006, \apj, 641, 689. doi:10.1086/500572
\bibitem[Wang et al.(2024a)]{wang24} Wang, B., de Graaff, A., Davies, R.~L., et al.\ 2024, arXiv:2403.02304. doi:10.48550/arXiv.2403.02304
\bibitem[Wang et al.(2024b)]{wang24b} Wang, B., Leja, J., de Graaff, A., et al.\ 2024, \apjl, 969, L13. doi:10.3847/2041-8213/ad55f7
\bibitem[Williams et al.(2024)]{williams24} Williams, C.~C., Alberts, S., Ji, Z., et al.\ 2024, \apj, 968, 34. doi:10.3847/1538-4357/ad3f17
\bibitem[Willott et al.(2010)]{willott10} Willott, C.~J., Albert, L., Arzoumanian, D., et al.\ 2010, \aj, 140, 2, 546. doi:10.1088/0004-6256/140/2/546
\bibitem[Yang et al.(2021)]{yang21} Yang, J., Wang, F., Fan, X., et al.\ 2021, \apj, 923, 262. doi:10.3847/1538-4357/ac2b32
\bibitem[Yang et al.(2023)]{yang23} Yang, J., Wang, F., Fan, X., et al.\ 2023, \apjl, 951, 1, L5. doi:10.3847/2041-8213/acc9c8
\bibitem[York et al.(2000)]{york00} York, D.~G., Adelman, J., Anderson, J.~E., et al.\ 2000, \aj, The Sloan Digital Sky Survey: Technical Summary, 120, 3, 1579. doi:10.1086/301513
\bibitem[Yue et al.(2024)]{yue24} Yue, M., Eilers, A.-C., Ananna, T.~T., et al.\ 2024, \apjl, 974, L26. doi:10.3847/2041-8213/ad7eba
\bibitem[Zakamska et al.(2016)]{zakamska16} Zakamska, N.~L., Hamann, F., P{\^a}ris, I., et al.\ 2016, \mnras, 459, 3144. doi:10.1093/mnras/stw718

\end{thebibliography}
\end{document}